\documentclass[12pt]{article}
\usepackage{graphicx}
\usepackage{epsfig}
\usepackage{epstopdf}
\usepackage{amsmath}
\usepackage{amsfonts}
\usepackage{amssymb}
\newcommand{\be}{\begin{equation}}
\newcommand{\beq}{\begin{equation}}
\newcommand{\eeq}{\end{equation}}
\newcommand{\eq}{\begin{equation}}
\newcommand{\ee}{\end{equation}}
\newcommand{\beqa}{\begin{eqnarray}}
\newcommand{\eeqa}{\end{eqnarray}}
\newcommand{\bea}{\begin{eqnarray}}
\newcommand{\eea}{\end{eqnarray}}

\newcommand{\mc}[1]{\mathcal{#1}}

\newcommand{\al}{{\alpha}}

\newcommand{\lp}{\left(}
\newcommand{\rp}{\right)}
\newcommand{\nn}{\nonumber}

\newcommand{\eg}{{\it e.g.,}\ }
\newcommand{\ie}{{\it i.e.,}\ }

\newcommand{\sr}{{\textsf{R}}}
\newcommand{\sm}{{\textsf{M}}}
\newcommand{\sj}{{\textsf{J}}}
\newcommand{\sa}{{\textsf{A}_\textsf{H}}}

\numberwithin{equation}{section}

\begin{document}

\thispagestyle{empty}

\begin{titlepage}

\begin{flushright}
{\small
}
\end{flushright}
\vskip 1cm

\centerline{\Large \bf Black Rings in (Anti)-deSitter space}

\vspace{1cm}

\centerline{
Marco M.~Caldarelli$^{a,}$\footnote{caldarelli@ub.edu},
Roberto Emparan$^{a,b,}$\footnote{emparan@ub.edu}
and Mar\'\i a J.~Rodr\'\i guez$^{a,}$\footnote{majo@ffn.ub.es}}

\vskip .5cm
\centerline{$^a$\em Departament de F{\'\i}sica Fonamental, Universitat de
Barcelona,}
\centerline{\em Marti i Franqu\`es 1, E-08028 Barcelona, Spain}
\bigskip
\centerline{$^b$\em Instituci\'o Catalana de Recerca i Estudis
Avan\c cats (ICREA),}
\centerline{\em Passeig Llu{\'\i}s Companys, 23, E-08010 Barcelona, Spain}

\vskip 1cm

\vskip 1cm

\begin{abstract}

\noindent We construct solutions for thin black rings in Anti-deSitter
and deSitter spacetimes using approximate methods. Black rings in AdS
exist with arbitrarily large radius and satisfy a bound $|J|\leq LM$,
which they saturate as their radius becomes infinitely large. For
angular momentum near the maximum, they have larger area than
rotating AdS black holes. Thin black rings also exist in deSitter
space, with rotation velocities varying between zero and a maximum, and
with a radius that is always strictly below the Hubble radius. Our
general analysis allows us to include black Saturns as well, which we
discuss briefly. We present a simple physical argument why
supersymmetric AdS black rings must not be expected: they do not possess
the necessary pressure to balance the AdS potential. We discuss the
possible existence or absence of `large AdS black rings' and their
implications for a dual hydrodynamic description. An analysis of the
physical properties of rotating AdS black holes is also included.

\end{abstract}

\end{titlepage}

\setcounter{footnote}{0}


\newpage

\section{Introduction}

The discovery of black rings \cite{Emparan:2001wn,Emparan:2006mm} and
other novel black holes in asymptotically flat higher-dimensional
gravity \cite{Emparan:2008eg} naturally prompts the question of whether
they admit counterparts in the presence of a cosmological constant. In
particular, their existence (or absence) in Anti-deSitter space is of
special interest for the possible implications in the context of the
AdS/CFT duality. However, in spite of attempts since early on, an exact
solution describing an (Anti-)deSitter black ring remains elusive.

Nevertheless, there appears no obvious physical reason why these
solutions should not be possible. Black rings represent states of
equilibrium between the tension and the centrifugal force of a rotating
circular black string. Putting a ring in Anti-deSitter space should have
the effect of increasing the gravitational centripetal pull on it, but,
at least within some parameter ranges, this can be plausibly balanced by
spinning the black ring faster. On the other hand, if we put the ring in
deSitter space, the cosmological expansion should act against the
tension, and so the required rotation should be smaller and possibly
reach zero. Thus, the failure to find any exact (Anti-)deSitter black
rings seems to be no more than a technical snag. The situation in
this respect is similar to that of asymptotically flat black rings in
$d\geq 6$, for which exact analytic solutions seem unlikely, but
nevertheless thin black rings have been constructed via approximate
methods in every dimension $d\geq 5$ \cite{Emparan:2007wm}.

It seems natural, then, to apply the methods of \cite{Emparan:2007wm} to
the approximate construction of cosmological black rings. This is the
subject of this paper, and our results prove the above conjectures
correct. Namely, thin black rings exist in AdS$_{d\geq 5}$ with
arbitrarily large radius, which spin comparatively faster than their
asymptotically flat counterparts; and black rings are also possible in
dS$_{d\geq 5}$ with rotation velocities that range between zero and a
maximum. An exact black ring solution in (A)dS$_5$ may still be found,
but we doubt that this is possible in $d\geq 6$.

By construction, our black rings, with horizon topology $S^1\times
S^{d-3}$, are {\em thin}, in the sense that their $S^{d-3}$-radius is
much smaller than both the $S^1$-radius {\em and} the cosmological
radius. These black rings have negative specific heat, and in the
AdS/CFT context belong in the same class as `small AdS black
holes'\footnote{Small in the sense that one of the horizon dimensions is smaller
than the AdS radius. All our solutions have regular horizons of finite
area in the second-derivative theory.}:
metastable excitations of the thermal dual CFT that do not admit a dual
hydrodynamic description. However, one may still wonder whether `large
AdS black rings' exist beyond the reach of our approximate methods, and
whether they could correspond to hydrodynamic phases of the thermal CFT
that have been overlooked so far. We shall discuss the
available evidence to explore this possiblity.

May black rings also exist in other backgrounds? More precisely, given a
stationary spacetime with a spatial $U(1)$ isometry, is it possible to
place a (thin) black ring along the direction of an orbit of the
isometry? We have made some headway in answering in the affirmative this
question for static backgrounds. The perturbation analysis of
\cite{Emparan:2007wm} can be easily extended to a wide class of static
backgrounds, for which we prove that a boosted neutral black string that
satisfies a suitable, easy to verify, equilibrium condition, remains
regular on and outside its horizon when bent into a circular shape.

The approach of \cite{Emparan:2007wm} can also be readily extended to
black rings charged under gauge fields. If we apply it to the
supersymmetric black rings of five-dimensional supergravity
\cite{Elvang:2004rt}, we find a simple, physical explanation for why
supersymmetric AdS$_5$ black rings have not turned up despite many
attempts to find them\footnote{For published attempts, see \eg
\cite{Acharya:2006as,Kunduri:2006uh,Kunduri:2007qy}.}:
supersymmetric black strings have zero pressure,
so if we bend them and place them in AdS, they can not counterbalance
the centripetal pull of the AdS potential. This argument fits nicely
with the analysis of \cite{Kunduri:2006uh,Kunduri:2007qy}, which concluded
from a study
of near-horizon solutions that supersymmetric black rings can not exist
without conical singularities on the plane of the ring.

The paper is organized as follows. Section \ref{sec:thinrings} describes
how to obtain the equilibrium condition for black rings in AdS, and
using this, proceeds to analyze their physical properties. The technical
details of the perturbation theory for curving a black string into a
black ring are deferred to appendices \ref{app:adapted} and
\ref{app:bendstring}, which actually deal with a much more general
situation than required for (A)dS spacetimes. Section \ref{sec:mpvsring}
compares the properties of black rings and rotating AdS black holes with
a single angular momentum. In section \ref{sec:mergers} we discuss a
plausible extension of the phases of thin black rings that we have
constructed, and then we examine whether large AdS black rings may exist
that could be described within the fluid-dynamical approach to rotating
black holes in AdS \cite{Bhattacharyya:2007vs,Bhattacharyya:2008ji}.
Section \ref{sec:dsrings} demonstrates how a simple application of our
methods allows to conclude that black rings, including static ones, are
possible in deSitter space, and how a class of black Saturns can be
easily obtained too. Section \ref{sec:nosusy} argues that thin
supersymmetric black rings should not be expected in AdS$_5$. Section
\ref{sec:conclusions} concludes with remarks on the outlook from the
paper. Besides the appendices mentioned above, appendix \ref{app:MPAdS}
contains a brief discussion of some general properties of rotating AdS
black holes with several angular momenta.

\paragraph{Notation and terminology.} Throughout this work we find convenient to
introduce dimensionless magnitudes, denoted by sans-serif fonts,
corresponding to quantities measured in units of the cosmological radius
$L$ or `cosmological mass' scale $L^{d-3}/G$. For instance, for the
$S^1$-radius, mass, angular momentum and horizon area of the ring we
define
\beq\label{dless}
\sr=\frac{R}{L}\,,\qquad \sm =\frac{GM}{L^{d-3}}\,,\qquad
\sj=\frac{GJ}{L^{d-2}}\,,\qquad \sa =\frac{A_H}{L^{d-2}}\,.
\eeq
Equivalently, we might have set $L=1=G$, but the meaning of some formulas is
clearer if we retain $L$.

To avoid misnomers, we shall refer to the
solutions that generalize the Kerr and Myers-Perry black holes to
include a cosmological constant
\cite{Carter:1968ks,Hawking:1998kw,Gibbons:2004uw} as {\em rotating
(A)dS black holes}. Even if (A)dS black rings and pinched black holes
are also (A)dS black holes, we hope that the distinction is clear from
the context.


\section{Thin black rings in AdS}
\label{sec:thinrings}

\subsection{The approximation}

Our analysis of thin black rings in Anti-deSitter space extends the
basic idea developed in \cite{Emparan:2007wm}. The rings, with horizon topology
$S^1\times S^{d-3}$, are characterized by the two radii $R$ and $r_0$ of the
$S^1$ and $S^{d-3}$ respectively. They are built by bending a straight,
thin, boosted black string into a circular shape.
At large distances from the horizon of the black ring, the latter is
approximated by a distributional source of energy-momentum, while near
the ring the solution is obtained by perturbing the straight black
string into an arc of large radius $R$.

In the present case, besides the circle
radius $R$ we have another scale from the Anti-deSitter curvature radius $L$,
which we also take to be much larger than $r_0$. It is clear that if
$R\ll L$ the solutions are only slight
modifications of the asymptotically flat black rings of \cite{Emparan:2007wm}.
However, we do not want
to assume any hierarchy among $R$ and $L$, in particular we want to
describe black rings for which $R>L$, which may behave
differently than those with $R\ll L$. In appendix \ref{app:adapted} we
show that our
approximations are in fact valid whenever\footnote{Actually we derive a
more precise form,
$r_0\ll R/\sqrt{1+\sr^2}$, equivalent to \eqref{smallroRL}
within factors of order one: this is the condition that $r_0$ be much
smaller than the extrinsic curvature radius of the $S^1$.}
\beq\label{smallroRL}
r_0\ll \min{(R,L)}\,,
\eeq	
which allows us to treat rings longer than the cosmological scale, $R>
L$, as long as the $S^{d-3}$ is small on that scale, $r_0<L$.

To be specific, consider global AdS$_d$ spacetime,
\eq
ds^2=-V(\rho)\,d\tau^2+\frac{d\rho^2}{V(\rho)}+\rho^2\lp d\Theta^2
+\sin^2\Theta\, d\Omega^2_{d-4}+\cos^2\Theta\,d\psi^2\rp,
\label{globalads}\eeq
where
\eq\label{Vads}
V(\rho)=1+\frac{\rho^2}{L^2}.
\eeq
We shall place the ring at $\rho=R$ on the
$\Theta=0$ plane. Observe that $R$ measures the proper length of the
circle. It is natural to define proper time and length coordinates, $t$
and $z$, along the worldsheet of the ring as
\beq\label{orthocoord}
t=\sqrt{V(R)}\;\tau=\sqrt{1+\sr^2}\;
\tau\,,\qquad
z= R\,\psi\,.
\eeq
The coordinate $z$ must be regarded as periodic, $z\sim z+2\pi R$.
Observe that the AdS background has the effect of introducing a
redshift between the proper time coordinate $t$ near the ring and the
canonically-normalized asymptotic time $\tau$.

At large distance in the directions transverse to the ring, the
gravitational field it creates is the same as that of an equivalent
circular distribution of energy-momentum, $T_{\tau\tau}$,
$T_{\tau\psi}$, $T_{\psi\psi}$, centered at $\rho=R$ on the $\Theta=0$
plane in \eqref{globalads}. We specify the source locally using the
orthonormal coordinates $(t,z)$ along the string worldsheet, in which it
takes the same form as
for a straight boosted string in flat space,
\begin{eqnarray}\label{tensorcomponents}
&&T_{tt}=\frac{r_0^{d-4}}{16\pi G}\lp (d-4)\cosh^2\al+1\rp\;\delta^{(d-2)}(r)\,,\nonumber\\
&&T_{t  z}=\frac{r_0^{d-4}}{16\pi G}\;(d-4)\cosh\al\sinh\al\;\delta^{(d-2)}(r)\,,\\
&&T_{z z}=\frac{r_0^{d-4}}{16\pi G}\lp (d-4)\sinh^2\al-1\rp\;\delta^{(d-2)}(r)\,,\nonumber
\end{eqnarray}
where $\al$ is the boost parameter along the
string and $r=0$ is the location of the ring\footnote{We normalize
$\int_{B^{d-2}}\delta^{(d-2)}(r)=\Omega_{d-3}$ for a ball $B^{d-2}$ that
intersects $r=0$ once.} (see
appendix~\ref{app:adapted} for the
construction of coordinates adapted to the ring).

\subsection{Equilibrium condition}

Not every energy-momentum source can be consistently coupled to gravity:
it must satisfy the conservation equations $\nabla_\mu T^\mu{}_\nu=0$.
For a thin brane, such as our black ring, this is equivalent to imposing
the `equations of motion' ${K_{\mu\nu}}^\rho T^{\mu\nu}=0$, derived by
Carter in \cite{Carter:2000wv}, where ${K_{\mu\nu}}^\rho$ is the
extrinsic curvature tensor of the brane embedding.

For a circular source at $\rho=R$, $\Theta=0$ in a metric of the generic
form \eqref{globalads}, these
equations imply
\beq
\frac{R V'(R)}{2V(R)}\,T^\tau{}_\tau+ T^\psi{}_\psi=0\,,
\label{equileq}\eeq
so in the case of Anti-deSitter \eqref{Vads}\footnote{The general case
for $V(\rho)$
will be studied in 
section~\ref{sec:dsrings}.},
\eq\label{equil}
\frac{\sr^2}{1+\sr^2}\,T^\tau{}_\tau+T^\psi{}_{\psi}=0\,.
\eeq
Observe that, whereas asymptotically flat black rings $(\sr=0)$ are
sourced by a pressureless distribution of matter, the attraction caused
by the AdS potential demands a non-vanishing pressure of the circular
source to achieve equilibrium.

Since $T^\tau{}_\tau=T^{t}{}_{t}=-T_{tt}$ and
$T^\psi{}_{\psi}=T^{z}{}_{z}=T_{zz}$, using \eqref{tensorcomponents} we
find that the equilibrium condition \eqref{equil} amounts to
\eq
\sinh^2\alpha=\frac{1+(d-2)\,\sr^2}{d-4}\,.
\label{boost}
\eeq
This condition fixes the value of the boost that provides the correct
centrifugal force to counterbalance the ring tension and the AdS
gravitational potential. As $\sr\to 0$ we recover the finite, flat-space
value obtained in \cite{Emparan:2007wm}. The cosmological constant has the effect that
the equilibrium boost depends on the ring radius: the longer the ring,
the stronger the cosmological pressure on it, so the boost increases with
$\sr$ and approaches the speed of light as $\sr\to\infty$.

Eq.~\eqref{boost} is a necessary condition for the existence of a
regular solution for a thin black ring. But in addition, we must also
construct the metric perturbation of the boosted black string that curves it
into a circular shape, and check that the perturbed horizon remains
regular. This calculation is described in appendix~\ref{app:bendstring}.

\subsection{Physical properties}

We compute the mass and angular momentum of the distributional ring as
\beq
M=\int_{\Sigma_\tau} T_{\mu\nu}n^\mu \xi^\nu \,,
\eeq
\beq
J=\int_{\Sigma_\tau} T_{\mu\nu}n^\mu \chi^\nu \,,
\eeq
where $\Sigma_\tau$ is a spatial section at constant time $\tau$, the vector
$n$ is the unit future-directed normal to $\Sigma_\tau$ and
$\xi=\partial_\tau$ and $\chi=\partial_\psi$ are the asymptotically
canonically normalized Killing vectors conjugate to the mass and angular
momentum. Since
\beq
T_{\mu\nu}n^\mu \xi^\nu=\sqrt{1+\sr^2}\, T_{tt}\,,\qquad
T_{\mu\nu}n^\mu \chi^\nu =R\, T_{t z}
\eeq
we find, substituting the equilibrium value \eqref{boost} into
\eqref{tensorcomponents},
\beq\label{ringM}
M=\frac{r^{d-4}_0\,L}{8
G}\Omega_{d-3}(d-2)\sr\left(1+\sr^2\right)^{3/2}\,,
\eeq
and
\beq\label{ringJ}
J=\frac{r^{d-4}_0\,L^2}{8
G}\Omega_{d-3}\sr^2 \left[
\left(1+(d-2)\sr^2\right)\left(d-3+(d-2)\sr^2\right)
\right]^{1/2}\,.
\eeq
(without loss of generality, we shall assume throughout the paper that
$J$ is positive). Thus
we obtain that the equilibrium of rings in AdS requires that the radius
be fixed in terms of the angular momentum per unit mass by
\beq\label{JonM}
\frac{J}{M L}=\frac{1}{d-2}\frac{\sr}{\left(1+\sr^2\right)^{3/2}}
\left[
\left(1+(d-2)\sr^2\right)\left(d-3+(d-2)\sr^2\right)
\right]^{1/2}\,.
\eeq

Other physical properties of the black ring are easily obtained. The
horizon area is the area of a boosted black string of length $\Delta z=2\pi R$,
\beq\label{ringAH}
A_H=2\pi R\, \Omega_{d-3} r_0^{d-3}\cosh\al=2\pi Lr_0^{d-3}\Omega_{d-3}
\sr\sqrt{\frac{d-3+(d-2)\sr^2}{d-4}}\,.
\eeq
The horizon of the boosted black string is generated by the Killing
vector
\beqa
\hat\zeta &=&\partial_{t}+\tanh\al\, \partial_{z}
=\frac{1}{\sqrt{1+\sr^2}}\left(
\partial_{\tau}+\frac{\sqrt{1+\sr^{2}}}{R}\tanh\al\,
\partial_{\psi} \right)\nonumber\\
&=&\frac{1}{\sqrt{1+\sr^2}}\zeta
\eeqa
where $\zeta=\partial_\tau+\Omega_H \partial_\psi$ is the null generator of the
horizon in terms of the Killing generators of time translations and
rotations at infinity. Thus we find the angular velocity of the horizon
\beq
\Omega_H=\frac{\sqrt{1+\sr^2}}{R}\tanh\alpha=
\frac{1}{L}
\sqrt{\frac{(1+\sr^{2})(1+(d-2)\sr^2)}{\sr^2(d-3+(d-2)\sr^2)}}\,.
\eeq
Observe that
\beq
\Omega_H >\frac{1}{L}\,,
\eeq
and $\Omega_H L\to 1$ as $\sr\to\infty$. For the rotating AdS black hole
solutions an angular velocity $\Omega_H L>1$ was argued in
\cite{Hawking:1999dp} to be associated with the existence of ergoregions
relative to asymptotic observers, which in AdS are believed to yield a
superradiant instability \cite{Cardoso:2004hs,Kunduri:2006qa}. Certainly
we expect our thin black rings to
possess these ergoregions\footnote{The non-superradiant ergoregion of
the boosted black string becomes superradiant when bent into a
ring-shape since the asymptotic behavior of modes changes.} and thus
they should suffer from this instability. However, the time scale of the
instability may be very long for thin rings. 

On the other hand, our black rings should suffer from
Gregory-Laflamme-type instabilities \cite{Gregory:1993vy} along the
direction of the ring. In fact the boost enhances the instability of
the black string by reducing the wavelength of the threshold mode,
$\lambda_{GL}\sim r_0/\cosh\alpha$ \cite{Hovdebo:2006jy}, so for a given
radius $R$ the black ring in AdS fits more easily the unstable
wavelengths than in the asymptotically flat case.

The surface gravity $\kappa$ of the Killing horizon generated by $\zeta$ is
related through the redshift factor to the surface gravity $\hat\kappa$
associated to
$\hat\zeta$,
\beq
\kappa=\sqrt{1+\sr^2}\,\hat\kappa=\sqrt{1+\sr^2}\frac{d-4}{2 r_0\cosh\alpha}=
\frac{(d-4)^{3/2}\sqrt{1+\sr^2}}{2
r_0\sqrt{d-3+(d-2)\sr^2}}\,.
\eeq

Our results for $M$, $J$, $A_H$, $\Omega_H$ and $\kappa$ are in
principle valid up to corrections of order $r_0/\min{(R,L)}$. For $d\geq
6$ they are very likely correct up to the next order: in the same way as
argued in \cite{Emparan:2007wm}, the dipole corrections to the metric
induced at order $r_0/\min{(R,L)}$ do not generate any corrections to
the physical magnitudes. The only possibility for corrections at this
order is through gauge modes becoming physical through boundary
conditions. In \cite{Emparan:2007wm}, these effects were present only in
$d=5$, and this is likely the case as well in AdS.

It is straightforward to check that the first law
\beq
dM=\frac{\kappa}{8\pi G}d A_H +\Omega_H dJ\,,
\eeq
is satisfied. However, the naive Smarr relation is not satisfied, since
\beq
\frac{d-3}{d-2}M\neq\frac{\kappa}{8\pi G} A_H +\Omega_H J\,.
\eeq
This is due to the presence of the cosmological length scale, which
invalidates the standard scaling argument that connects the first law
and the Smarr relation.

A ``BPS bound" on the angular momentum of regular solutions in AdS was
proven in \cite{Chrusciel:2006zs}, which, for solutions with a single
(positive) angular
momentum, takes the form
\beq
J\leq ML\,.
\eeq
It is easy to check that this bound is satisfied by our black rings.
It is also saturated in the limit of very long rings, $\sr\to\infty$,
where the difference between $M L$ and $J$ at fixed mass decreases
to zero as $\sr\to\infty$,
\beq
ML -J=\frac{ML}{\sr^2}+\mc{O}(\sr^{-4})\,.
\eeq

For a given $L$, the solutions are characterized by the dimensionless mass
$\sm$ and angular momentum $\sj$ defined in \eqref{dless}.
The BPS bound is equivalently expressed as $\sj\leq \sm$. For fixed mass
$\sm$, the horizon area decreases to zero as the bound is approached like
\beq\label{amaxjr}
\sa \propto \sm^{\frac{d-3}{d-4}}\left(1-\frac{\sj}{\sm}\right)^{\frac{d-2}{d-4}}
\left(1+\mc{O}(\sm-\sj)\right)\,
\eeq
(we omit the numerical proportionality factor, which we will not require).

\section{Black rings vs.\ rotating AdS black holes}
\label{sec:mpvsring}

In order to make a comparison to rotating AdS black holes, we first
analyze the properties of the latter. The solutions were found in
\cite{Hawking:1998kw,Gibbons:2004uw} and their physical parameters
correctly computed in \cite{Gibbons:2004ai}. We consider the situation
with only one spin turned on\footnote{Some properties of solutions with
several spins are discussed in appendix \ref{app:MPAdS}.}, for which the
solutions are parametrized by the mass and rotation parameters $m$ and
$a$, and the physical magnitudes are
\beq\label{MPAmass}
M=\frac{\Omega_{d-2}}{4\pi G}\frac{m}{\Xi^2}\left(1+\frac{(d-4)\Xi}{2}\right)\,,
\eeq
\beq\label{MPAspin}
J=\frac{\Omega_{d-2}}{4\pi G}\frac{ma}{\Xi^2}\,,
\eeq
\beq
A_H=\Omega_{d-2} r_+^{d-4}\frac{r_+^2+a^2}{\Xi}\,,
\eeq
\beq
\Omega_H=\frac{a}{L^2}\frac{r_+^2+L^2}{r_+^2+a^2}\,,
\eeq
\beq
\kappa=r_+\left(1+\frac{r_+^2}{L^2}\right)\left(\frac{d-3}{2
r_+^2}+\frac{1}{r_+^2+a^2}\right)-\frac{1}{r_+}\,,
\eeq
with
\beq
\Xi\equiv1-\frac{a^2}{L^2}
\eeq
and $r_+$ being the largest positive real root of
\beq\label{rplus}
r_+^{d-5}(r_+^2+L^2)(r_+^2+a^2)=2m L^2\,.
\eeq
The rotation parameter is restricted to $a\leq L$.
It is then easy to see that these solutions satisfy the bound $J\leq
ML$, with $J/ML\to 1$ as $a\to L$.

\subsection{Range of angular momentum}

There are significant differences
between $d=5$ and $d\geq 6$ in how the
spin is bounded above at any given mass, and in the nature of the
solutions at maximum angular momentum. These are the AdS counterparts of
properties of the asymptotically flat case. In many respects, the limit
$a\to L$ and the bound $\sj=\sm$ in AdS correspond to the limit
$a\to\infty$ and the ultra-spinning regime of infinite spin for fixed
mass of Myers-Perry (MP) black holes \cite{Myers:1986un,Emparan:2003sy}.

In $d=5$ eq.~\eqref{rplus} can be solved explicitly. Real values of
$r_+$ that yield black hole solutions
exist provided
that $a^2< 2m$. This
implies that we cannot take $a\to L$ and at the same time keep the mass
$M$ fixed at a finite value, so the BPS bound
$\sj=\sm$ is never saturated. In the limit $a^2\to 2m$ we have $r_+\to
0$ and the surface gravity $\kappa\to 0$: this is an extremal naked
singularity of zero area, with maximum angular momentum for given
mass,
\beq\label{jmax5d}
\sj_{max}^2(\sm)=\sm^2+\frac{9\pi}{16}\sm+\frac{27\pi^2}{512}
-\sqrt{\pi}\left(\sm+\frac{9\pi}{64}\right)^{3/2}<\sm^2\,\qquad (d=5)\,.
\eeq
At small $\sm$ this reduces to the known asymptotically flat extremality
bound, $\sj^2_{max}\to 32 \sm^3/27\pi$, while at very large $\sm$ the extremal
angular momentum approaches the BPS bound,
$\sj_{max}\to\sm(1-\sqrt{\frac{\pi}{4\sm}})$.

\begin{figure}[th]
\centerline{\includegraphics[width=16.5cm]{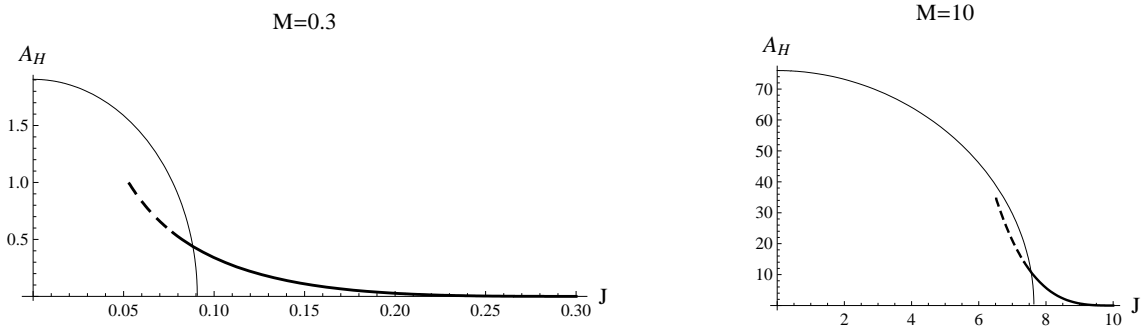}}
\caption{\small Plots in five dimensions of $\sa(\sj)$ for fixed $\sm$,
at small $\sm$ (left) and large $\sm$ (right). Thin lines correspond to
rotating AdS black holes, thick lines to black rings. The thick solid
line continues to a thick dashed line when the thin-ring approximation
$r_0\ll R/\sqrt{1+\sr^2}$ breaks down (here and in the next figures the
solid-dashed divide is arbitrarily taken at
$r_0=\frac{R}{5\sqrt{1+\sr^2}}$ and the dashed line is extended up to $r_0=
\frac{R}{2\sqrt{1+\sr^2}}$). The spin of black rings reaches up to the
BPS bound $\sj=\sm$, but the spin of rotating AdS black holes is bounded by
$\sj\leq \sj_{max}(\sm)<\sm$, eq.~\eqref{jmax5d}. For small $\sm$, and
$\sj$ not too close to $\sm$, the curves are very similar to the
asymptotically flat case.}
\label{fig:AJ5D}
\end{figure}

In $d\geq 6$, instead, the same argument of the asymptotically flat
case \cite{Myers:1986un} shows that for any value of
$a\in[0,L)$ and for finite mass there
is always
a real positive solution
to \eqref{rplus}. In the limit $a\to L$, which requires taking $m\to 0$
in order to keep the mass finite,
we find
\beq
\sj_{max}(\sm)=\sm \qquad (d\geq 6)\,,
\eeq
so the BPS bound is saturated.
In this limit one finds $r_+\to 0$.
However, the solutions that saturate the bound are not extremal, in
the sense that the surface gravity does not vanish, but instead
diverges like $\kappa\to (d-5)/2r_+$. The area vanishes in the limit,
decreasing to zero like
\beq\label{amaxjh}
\sa \propto \sm^{\frac{d-4}{d-5}}\left(1-\frac{\sj}{\sm}\right)^{\frac{d-3}
{d-5}}\left(1+\mc{O}(\sm-\sj)\right)\,.
\eeq
Comparing to \eqref{amaxjr}, we see that near the maximum $\sj$, black
rings always have larger area than rotating AdS black holes with the same
mass.

\begin{figure}[th]
\centerline{\includegraphics[width=17cm]{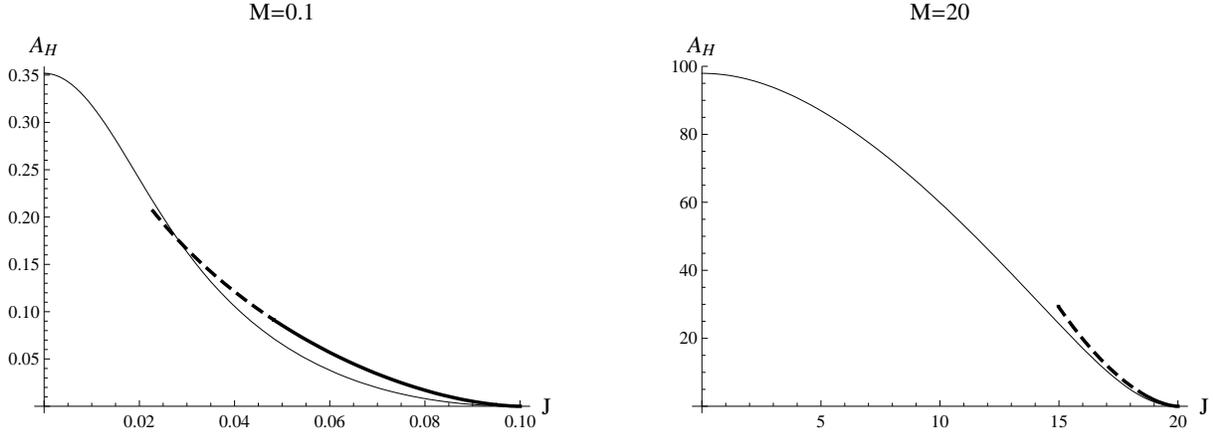}}
\caption{\small Plots in seven dimensions of $\sa(\sj)$ for fixed $\sm$,
at small $\sm$ (left) and large $\sm$ (right). Thin lines correspond to
rotating AdS black holes, thick lines to black rings (dashed
line when the thin-ring approximation breaks down). Both rotating AdS black
holes and black rings extend up to the BPS bound $\sj=\sm$, but black
rings have larger area near the maximum $\sj$. The asymptotic curves
near this point are \eqref{amaxjr} and \eqref{amaxjh}.}
\label{fig:AJ7D}
\end{figure}

We plot the curves $\sa(\sj)$ for fixed $\sm$, at small and large $\sm$,
in five dimensions, fig.~\ref{fig:AJ5D}, and in seven dimensions,
fig.~\ref{fig:AJ7D}, as a representative of
$d\geq 6$.

\subsection{Limits to black membranes.}

There are other features that justify referring to the limit $a\to L$ in
$d\geq 6$ as an analogue of the `ultraspinning limit' of asymptotically
flat MP black holes. Here we follow \cite{Emparan:2003sy}. The sectional
areas of the black hole in the directions transverse and parallel to the
rotation plane allow to identify a transverse and parallel size of the
horizon,
\beq\label{ells}
\ell_\perp\sim r_+\,,\qquad
\ell_\| \sim \left(\frac{r_+^2+a^2}{\Xi}\right)^{1/2}\,
\qquad (a\to L)\,
\eeq
(it is slightly convenient to use $\Xi$ as the small parameter). In this
limit $r_+$ decreases like $\sim
L\,\sm^{\frac{1}{d-5}}\,\Xi^{\frac{2}{d-5}}$ so at fixed mass the ratio
$\ell_\|/\ell_\perp$ diverges like $\sim \Xi^{-\frac{d-1}{2(d-5)}}$: the
horizon pancakes out along the plane of
rotation. The geometry in fact
approaches that of a black membrane as the upper bound on $\sj$ is
approached. To see this, take the solution in Boyer-Lindquist
coordinates $(t,r,\theta,\phi,\Omega_{d-4})$ of \cite{Gibbons:2004uw},
and define a new mass parameter and new coordinates
\beq\label{memlimit}
\hat \mu=\frac{2m}{L^2\Xi^2}\,,\qquad \hat t= \Xi^{-\frac{2}{d-5}}t\,,
\qquad \hat r=\Xi^{-\frac{2}{d-5}}r\,,\qquad
\sigma=\Xi^{-\frac{d-7}{2(d-5)}}L\sin\theta
\eeq
that remain finite as $a\to L$, \ie as $\Xi\to 0$.
In the limit the metric becomes
\beq\label{memmet}
ds^2\to \Xi^{\frac{4}{d-5}}\left[-\left(1-\frac{\hat\mu}{\hat
r^{d-5}}\right)d\hat t^2
+\frac{d\hat r^2}{1-\hat\mu/\hat r^{d-5}}+d\sigma^2+\sigma^2
d\phi^2+\hat r^2 d\Omega^2_{d-4}
\right]\,.
\eeq
This is the metric of a black membrane, only rescaled by a conformal
factor that reflects the fact that if we keep the mass of the black
hole fixed, then its size goes to zero in the limit $a\to L$. The
metric \eqref{memmet} should give a good
approximation to the horizon geometry up to $\sigma\sim
L\,\Xi^{-\frac{d-1}{2(d-5)}}$.

Like in \cite{Emparan:2003sy}, it is natural to conjecture that a
Gregory-Laflamme-type of instability appears for black holes
that approach this limit. For a fixed value of $\sm$, the onset of the
black membrane-like behavior of the black holes can be located at the
value of $\sj$ for which the surface gravity $\kappa$ reaches a minimum,
which is also an inflection point for $\sa$. One can find the critical
value of $r_+/L$ for which the minimum of $\kappa$ occurs at a given
mass. It turns out that this critical value $r_{+c}$ is always in a
range $r_{+ c}\leq\sqrt{\frac{d-5}{d-1}} L$, approaching the latter
value for large mass. The corresponding critical spin $\sj_c$ is at
$(\sm-\sj_c)/\sm\sim \sm^{-1/2}$, very
near the maximum spin. Since the unstable regime takes place
at values $r_+<r_{+ c}$, we see that the black hole only becomes
unstable when its size $\ell_\perp$ is smaller than the AdS radius. I.e.,
large AdS black holes, with $\ell_\|,\ell_\perp >L$, should not suffer
from Gregory-Laflamme-type instabilities.

It is worth discussing a limit of the rotating AdS solutions that does {\em
not} have an analogue in the absence of the cosmological constant,
so it necessarily belongs in the regime of very large $\sm$. In the
limit \eqref{memlimit} we kept the mass fixed, with the effect that the
black hole size decreased to zero as $a\to L$. Instead, if we insist on
keeping the horizon size finite, then as $a\to L$ we only scale
the polar angle $\theta\to 0$ while keeping finite a new coordinate $\sigma$
\beq
\sin\theta=\sqrt{\Xi} \, \sinh(\sigma/2)\,.
\eeq
The limiting metric is
\beq\label{hypermem}
ds^2=-f(r)\left(dt-L\sinh^2(\sigma/2)d\phi\right)^2
+\frac{dr^2}{f(r)}+\frac{L^2}{4}\left(1+\frac{r^2}{L^2}\right)
\left(d\sigma^2+\sinh^2\sigma d\phi^2\right)+r^2 d\Omega^2_{d-4}\,,
\eeq
where
\beq
f(r)=1-\frac{2m}{r^{d-5}(r^2+L^2)}+\frac{r^2}{L^2}\,.
\eeq
This solution (a particular case of solutions in
\cite{Mann:2003zh,Mann:2005ra}) has a horizon at the
largest real root $r=r_+$ of $f(r)$. The horizon geometry is
$\mathbb{H}^2\times S^{d-4}$, where the hyperboloid $\mathbb{H}^2$ has
coordinates $(\sigma,\phi)$. The geometry may then be regarded as a {\em
rotating black hyperboloid membrane}\footnote{Distinct from hyperbolic
AdS black holes \cite{Birmingham:1998nr}, whose horizon is
$\mathbb{H}^{d-2}$. It is also different than the conventional limit of
very large mass in which one recovers a static black hole with planar
horizon $\mathbb{R}^{d-2}$.}. The boundary
geometry is not the Einstein universe $\mathbb{R}_t\times S^{d-2}$ that
appears at finite mass, but rather a rotating spacetime with spatial
sections $\mathbb{H}^2\times S^{d-4}$.
Observe that the solution makes sense in
$d=5$ as well, where it will have a horizon provided that $m>L^2/2$. In
this case the solution, with horizon $\mathbb{H}^2\times S^1$, can also
be regarded as a rotating hyperbolic black string. Static hyperbolic black
strings were obtained in \cite{Mann:2006yi}.

The mass and angular momentum of the black hole diverge in this limit:
since we are keeping $m$ finite, $M$ and $J$ in \eqref{MPAmass},
\eqref{MPAspin} diverge as $a\to L$. However, their ratio remains
finite, $J/ML\to 1$. The hyperboloid membrane must then be interpreted
as the limit of very large $\sm$ of the black holes with spins very near
the maximum $\sj_{max}$. For small $r_+/L$ (which only happens in $d\geq
6$ for small $m$) the $\mathbb{H}^2$ radius is $\sim L$, much larger
than the $S^{d-4}$ size $\sim r_+$. In fact, when $r_+$ is very small
the geometry near the axis $\sigma=0$ reduces to the black membrane of
\eqref{memmet}. Presumably we should expect a Gregory-Laflamme
instability in this case. When $r_+$ is large the characteristic radii
of the hyperboloid and the sphere are both $\sim r_+$ and a GL
instability seems unlikely\footnote{Ref.~\cite{Brihaye:2007ju} studied
the GL instability of static hyperbolic black strings, and found them to
be stable. In this case the perturbations consisted of ripples along the
$S^1$, not the $\mathbb{H}^2$. But in any case, our argument suggests
that in $d=5$ the latter perturbations should also be stable.}.

\bigskip

We can summarize the results of this section as follows: the properties
of rotating AdS black holes at any given mass are qualitatively similar to
those of their asymptotically flat counterparts, only with a phase
diagram where the infinite range of angular momenta for MP black holes
is `compressed' to the range $\sj\leq \sm$ for rotating AdS black hole solutions. The
onset of the membrane behavior, and accompanying instabilities, happens
at values of the spin that, for small mass, are similar to the
asymptotically flat ones, but for large mass this region is pushed close
to the upper end of the phase space, $(\sm-\sj)/\sm\sim \sm^{-1/2}$. In
this regime, the differences to the asymptotically flat case that persist as
the mass is scaled up are captured by the limiting geometry
\eqref{hypermem}.

\section{Phase connections}
\label{sec:mergers}

Given the conclusion of the previous section, it seems reasonable to
conjecture that the curves for black rings in the phase diagrams at fixed
mass in figs.~\ref{fig:AJ5D} and \ref{fig:AJ7D} are completed in a
manner analogous to that proposed in \cite{Emparan:2007wm}, only
compressed to a range of $\sj\in [0,\sm]$. However, while the pattern of
connections proposed in \cite{Emparan:2007wm} can presumably be
imported without much distortion to AdS for small $\sm$, we cannot
expect it to provide more than a qualitative picture of the mergers
between phases when $\sm$ is large and the cosmological length scale
plays an important role.

\subsection{The girths of a black ring}

Black rings in AdS are characterized by the relative sizes of three
scales, $r_0$, $R$, and $L$, \ie the radius of the ring's $S^2$, $S^1$,
and of AdS, respectively. So far we have referred to rings
that satisfy \eqref{smallroRL} as thin, but for the purpose of the
following discussion it is convenient to introduce a more refined
terminology:
\begin{itemize}
\item {\em Thin} rings have
$r_0\ll R$; {\em fat} rings have $r_0\sim R$.

\item {\em Small} rings have $r_0<L$; {\em large} rings have  $r_0>L$.

\item {\em Short} rings have $R<L$; {\em long} rings have $R>L$.

\end{itemize}

Thus our approximation \eqref{smallroRL} applies to thin small rings,
either short or long. The definition of $r_0$ and $R$ for fat rings may
not be unambiguous, but we use the characterization $r_0\sim R$ to mean
that the self-gravitational attraction of the ring is very strong and
large curvature may be expected on the horizon, in particular near the
inner circle of the ring. With a suitable definition, $r_0$ will never
be larger than $R$, so not all combinations of girths are possible.
The same characterization could be used for rotating AdS black holes
if in place of $r_0$ and $R$ we use the sizes of the horizon in the
directions transverse and parallel to the rotation plane.

We are interested in the different possibilities for black rings at a
given value of the mass $\sm$. To begin with, we investigate the range of $\sm$
that the approximation \eqref{smallroRL} allows to cover. If we
consider rings at small $\sr$, for which \eqref{ringM} reduces to
\beq
\sm \sim \left(\frac{r_0}{L}\right)^{d-4} \sr
\eeq
then \eqref{smallroRL} implies that our approximation is only valid for
very small masses
\beq
\sm\ll \sr^{d-3}\qquad (\sr\ll 1)\,.
\eeq
This is actually the same condition as in the asymptotically flat case,
$GM\ll R^{d-3}$, and is hardly surprising:
at small $\sr$ the ring is much smaller and shorter
than $L$ and hence must be light on the cosmological scale. However, at
large $\sr$  we find from \eqref{ringM} and
\eqref{smallroRL} that we must require
\beq
\sm \sim \left(\frac{r_0}{L}\right)^{d-4} \sr^4 \ll \sr^{4}\qquad (\sr\gg 1)\,.
\eeq
Thus within our approximations we can get to the regime of large
$\sm$ for extremely long rings.

\subsection{The issue of large AdS black rings and
dual fluids}

Let us now explore heuristically how these branches of solutions may
extend when the approximation \eqref{smallroRL} breaks down. It seems
reasonable to assume that the curves of black rings, which we have seen
extend up to $\sj=\sm$, can only end by merging to a black hole of
spherical topology (possibly an extremal singular one in $d=5$, and a
pinched black hole in $d\geq 6$), and that this requires the ring to
become fat, $r_0\sim R$: the merger happens when the central hole of the
ring closes off.

\begin{figure}[t!]
\centerline{\includegraphics[width=11cm]{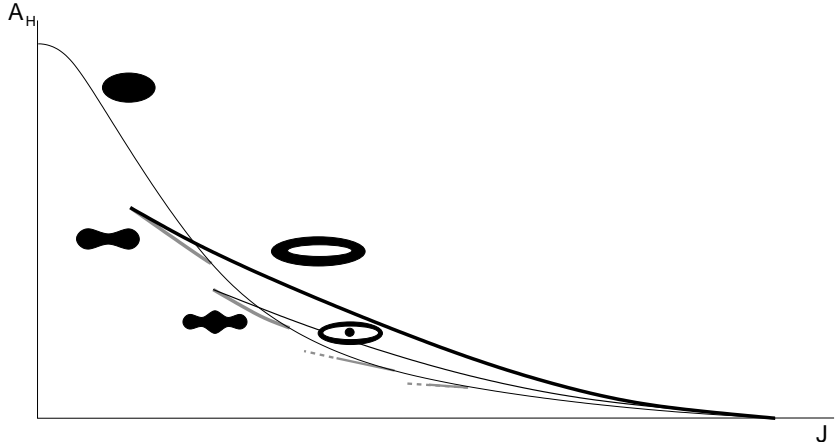}}
\caption{\small Proposal for the completion of phase curves at small $\sm$
in $d\geq 6$ (the situation in $d=5$ is a simple adaptation of the same
idea). The patterns proposed in \cite{Emparan:2007wm} are compressed
here to the range $\sj\leq \sm$. We stress that the details of the
connections (\eg first order vs.\ second order transitions) remain
unknown and are arbitrarily drawn.}
\label{fig:ads_loM}
\end{figure}

For very small $\sm$ the rings are all small and short, $r_0<R\ll L$.
This situation is very similar to the asymptotically flat case, in which
our approximation breaks down when the ring becomes fat, $r_0\sim R\ll
L$. Fat rings then connect to topologically spherical black holes or,
possibly in $d=5$, to a singular extremal solution. The scale $L$
introduces only small corrections. Fig.~\ref{fig:ads_loM} illustrates qualitatively
this situation in $d\geq 6$.

The regime of very large $\sm$ is more intriguing. In this case, we have
seen that our approximate methods reproduce some of these
solutions as very long, small black rings, $r_0\ll L\ll R$, at least
when $R\gg \sm^{1/4}L$. These cover the range of spins very close to the
upper bound $\sj \simeq \sm$. As we move away from this point, reducing
$\sj$ while keeping $\sm$ fixed, we are increasing $r_0$ and decreasing
$R$. At some point, if $\sm$ is very large, we must encounter a regime
where $L\sim r_0\ll R$ (increasing $r_0$ and decreasing $R$
cannot lead us to $R<L$ since these small short rings would not have
large $\sm$). It is unclear, however, whether $r_0$ can grow much larger
than $L$ while remaining much smaller than $R$, leading to a regime of
thin large rings.

This would appear problematic for two reasons. The first one concerns
the point at which black rings merge with black holes. In the previous
section we have observed that in $d\geq 6$, rotating AdS black holes at large
$\sm$ develop a membrane-like behavior at a radius $r_+\sim L$. It is
natural to expect that the threshold of a Gregory-Laflamme instability
sets in at around this point, giving rise to a branch of pinched black
holes. If these then connect to (fat) black rings with parameters in a
similar range, then the correspondence $r_+\sim r_0$ would imply that
these black rings would not be large. Unless the curve of AdS black
rings extends far from this merger point, thin large black rings may not
be realized, as $r_0$ would remain of the same order as $L$. Admittedly,
though, this argument is merely suggestive. 

\begin{figure}[t!]
\centerline{\includegraphics[width=16cm]{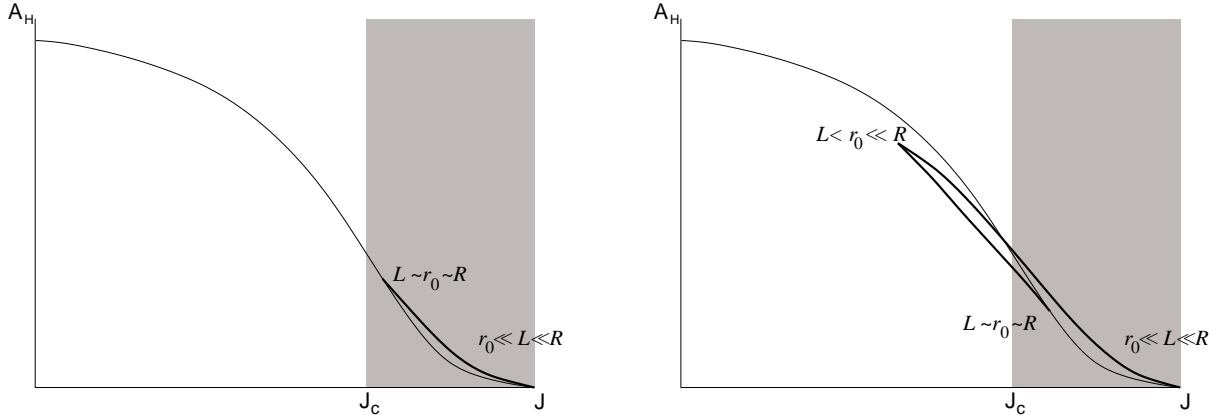}}
\caption{\small Two alternatives for the completion of phase curves at
$\sm\gg 1$. The gray-shaded area, which extends from $\sj
\sim \sj_c$ where $(\sm-\sj_c)/\sm\sim \sm^{-1/2}$, to the upper limit
$\sj=\sm$, corresponds to solutions that do {\em not} qualify as large
AdS black holes or rings. In the left picture, no large AdS black rings
exist. In the right picture, thin large AdS black rings, with $L<r_0\ll
R$ exist. This second alternative requires an explanation of why
these black rings do not appear in dual hydrodynamic studies.}
\label{fig:ads_hiM}
\end{figure}

The second concern is that, if black rings exist with parameters $L\ll
r_0\ll R$ then naively they might admit a dual fluid-dynamical
description: their curvature would remain smaller than $1/L$ and their
temperature larger than $1/L$. These conditions are required for the
applicability of the fluid-dynamical approximation (fat large black
rings, instead, probably cannot be described as dual fluids, as they
involve large curvatures). However, a thorough analysis of the
relativistic Navier-Stokes equation for a rotating fluid on the boundary
geometry $\mathbb{R}\times S^{d-2}$ has not revealed other solutions
than the duals of rotating AdS black holes \cite{Bhattacharyya:2007vs}
(see also \cite{Bhattacharyya:2008ji}). 

The apparent absence of thin large AdS black rings may not be too
striking. When $r_0\sim L$ the ring's $S^{d-3}$ may be highly distorted:
the ring is very massive and due to the centripetal cosmological pull,
it may `spread down towards the center of AdS' and thus become fat.
Since thin large rotating AdS black holes do exist which are highly
pancaked ---just take $r_+>L$ in \eqref{ells}, which makes $\sm \gg
1$--- one may naively envisage `drilling a hole' through them and
spinning them up faster to get a black ring. However, this may not be
possible since these black holes (which limit to the geometry
\eqref{hypermem}) are already very close to the maximum spin. 

A more quantitative argument that points to difficulties for a
regime of black rings with $L\ll r_0$ follows by first observing that,
quite naturally, $r_0$ should not be larger than the distance from the
ring to the center of AdS (otherwise the ring would close its hole). To
estimate this upper limit on $r_0$ for a given $R\gg L$, we can compute
the proper radial distance to $\rho=R$ in empty AdS space
\eqref{globalads}, namely
\beq
r_0^{max} \sim \int_{0}^{R}\sqrt{g_{\rho\rho}}\;d\rho =
\int_{0}^{R}\frac{d\rho}{\sqrt{1+\rho^2/L^2}}\simeq L\log(R/L)\,.
\eeq	
Although this estimate neglects possible strong curvature effects of the
ring, it suggests that the geometry of AdS severely limits
$r_0/L$ to grow only logarithmically with $R$, instead of the linear
growth in flat space. Thus, there may not exist a
regime proper of thin large AdS rings. Fig.~\ref{fig:ads_hiM}
illustrates the alternatives.

In conclusion, although the arguments are clearly not compellingly
strong, the absence of thin large black rings seems plausible, even for
very large $\sm$. If nevertheless they existed, then they should
presumably extend to values of $\sj$ in a range $(\sm-\sj)/\sm\gg
\sm^{-1/2}$ outside the regime of small long rings, possibly resembling
large rotating AdS black holes with a central hole. It would remain to
explain why they do not show up in the fluid-dynamical analysis of
\cite{Bhattacharyya:2007vs,Bhattacharyya:2008ji}. A likely possibility
is that for a black ring not all of the horizon is close to the boundary
(the inner rim of the central hole could lie deeper inside AdS), and
hence the fluid description could not capture all of it. In the dual
parlance, the fluid would become too thin near the poles of the
$S^{d-2}$ to be captured within the approximations of
\cite{Bhattacharyya:2007vs,Bhattacharyya:2008ji}. Or maybe there is a
subtlety missing in the analysis. But at the moment, the simplest
possibility is that the regime of thin large AdS rings is never
realized.


\section{Black rings galore}
\label{sec:dsrings}

\subsection{Black rings in deSitter}

The extension of our analysis to black rings in deSitter is straightforward:
just continue analytically the results for Anti-deSitter, $L\to iL$.
Thus, considering a metric of the form
\eqref{globalads} with the lapse function
\eq
V(\rho)=1-\frac{\rho^2}{L^2}\,,
\eeq
equilibrium fixes the boost parameter to
\eq
\sinh^2\alpha=\frac{1-(d-2)\,\sr^2}{d-4}\,.
\label{dsboost}\eeq
This is a decreasing function of $\sr$ that vanishes for
a maximum radius $\sr=\sr_{st}$,
\eq
\sr_{st}=\frac1{\sqrt{d-2}}\,.
\label{staticr}\eeq
For this value of the radius, we find a static thin black ring in
deSitter spacetime: the cosmological expansion exactly balances the
string tension. For $\sr>\sr_{st}$ the equilibrium condition cannot be
satisfied: these rings cannot avoid getting inflated away. Hence there
is an upper limit to the size of thin black rings in deSitter. Observe
that this upper limit is $\sr_{st}<1$. Even if it is clear that we
could not have super-Hubble rings with $\sr>1$, our thin
rings do not even approach this limit. However, it may
still be that fat black rings of larger size exist in deSitter, and they
might even reach the size of the cosmological horizon.

The existence of black rings in \textit{four-dimensional} deSitter space
is more difficult to rule out than in asymptotically flat or
Anti-deSitter spacetimes, where they are forbidden by topological
censorship theorems \cite{Galloway:1999bp}. It is clear that with our
methods we can not construct them, since the requisite vacuum
four-dimensional black strings do not exist. What this implies is that,
if black rings with horizon topology $S^1\times S^1$ existed at all in
four-dimensional deSitter, they could not admit a limit in which any of
the ring's $S^1$ became much smaller than the cosmological radius, \ie
they should always be \textit{fat}. They could neither be type-D
solutions \cite{Plebanski:1976gy}, so we find them very unlikely to
exist.

\subsection{Black Saturns}

Using our general formalism, we can easily study black rings living in
any static spherically symmetric spacetime. Appendices \ref{app:adapted}
and \ref{app:bendstring} show that neutral black strings can be placed
at an equatorial circle in these spacetimes to yield thin neutral black
rings that are regular on and outside the horizon.

A large class of physically interesting backgrounds is given by the
metric (\ref{globalads}), where $V(\rho)$ is an arbitrary function of
the radial coordinate. For example, with a proper choice of the function
$V(\rho)$ this allows to describe the bending of a black string in the
background of static black holes and thus construct thin black Saturns,
in the sense that the black ring surrounding the black hole is thin.
Exact solutions for black Saturns in five-dimensional vacuum gravity
were obtained in \cite{Elvang:2007rd}. Our approach here allows to
extend them and could be used to refine the arguments in
\cite{Elvang:2007hg}.

The rescaling encoding the redshift of the black string,
and the proper length along the string, are given by
\eq
t=\sqrt{V(R)}\tau\,,\qquad
z=R\psi\,.
\label{genredshift}\eeq

Using eqs.~\eqref{CV} and \eqref{genequil} we find that a circular
source with energy-momentum (\ref{tensorcomponents})
centered at $\rho=R$ on the $\Theta=0$ plane satisfies the equilibrium
condition (\ref{equileq}) as long as the boost parameter $\al$ of the
string verifies
\eq
\sinh^2\al=\frac1{d-4}\frac{2V(R)+(d-3)RV'(R)}{2V(R)-RV'(R)}\,.
\label{boostV}\eeq
Taking into account the redshift factors coming from the
rescalings (\ref{genredshift}), we can compute the physical quantities
characterizing the black ring in this general background. Its mass and
angular momentum read
\eq
M=\frac{r_0^{d-4}}{8G}\Omega_{d-3}R\,\frac{2(d-2)V^{3/2}}{2V-RV'}\,,
\eeq
\eq
J=\frac{r_0^{d-4}}{8G}\Omega_{d-3}R^2\,
\frac{\left[\lp2(d-3)V+RV'\rp\lp2V+(d-3)RV'\rp\right]^{1/2}}{2V-RV'}\,,
\eeq
where $V\equiv V(R)$ and $V'\equiv V'(R)$.
These $M$ and $J$ should be understood as quantities conjugate to the time $\tau$
and angular coordinate $\psi$ ---for each particular spacetime one has to
determine how these are related to the canonical choice of time, if the
spacetime admits one. For asymptotically flat or
asymptotically AdS solutions, the results above are the canonical ones.

The area is given by
\eq
A_H=2\pi r_0^{d-3}\Omega_{d-3}R\sqrt{\frac{2(d-3)V+RV'}{(d-4)(2V-RV')}}\,,
\eeq 
while the angular velocity, defined using the horizon generator
$\partial_\tau+\Omega_H\partial_\psi$, is
\beq
\Omega_H=\frac{\sqrt{V}}R\sqrt{\frac{2V+(d-3)RV'}{2(d-3)V+RV'}}
\eeq
and the surface gravity, for the same Killing vector, is
\eq
\kappa=\frac{(d-4)^{3/2}}{2r_0}\sqrt{\frac{(2V-RV')V}{2(d-3)V+RV'}}\,.
\eeq

For instance, choosing
\eq\label{schads}
V(\rho)=1-\frac{2m}{\rho^{d-3}}\pm\frac{\rho^2}{L^2},
\eeq
the above formulas yield the physical properties of thin
(Anti-)deSitter black Saturns with a static central black hole. Note
that in this approximation, the total mass of the system is the sum of
the mass of the central Schwarzschild-(A)dS black hole and the mass $M$
of the thin black ring surrounding it, while its total angular momentum
is given by $J$. 

Observe that there are two kinds of critical values of the ring radius:
the values $R=R_{l}$,
where
\beq
\frac{V'(R_{l})}{V(R_{l})}=\frac{2}{R_{l}}
\eeq
for which the boost \eqref{boostV} reaches speed of light (this is in
fact the radius of null circular orbits) so $r_0\to 0$ is
required if the mass, spin, and area remain finite; and the values
$R=R_{st}$ where the boost vanishes
\beq
\frac{V'(R_{st})}{V(R_{st})}=-\frac{2}{(d-3)R_{st}}
\eeq
and so the black ring is static.
The set of these critical
radii
determine the ranges of existence of black rings in these backgrounds.
We have seen one example in deSitter, eq.~\eqref{staticr}. For a black
Saturn in (A)dS, described by \eqref{schads}, the speed-of-light radius
\beq
R_{l}^{d-3}=(d-1)m=
\frac{d-1}{2}\rho_{bh}^{d-3}\left(1\pm\frac{\rho_{bh}^2}{L^2}\right)
\eeq
sets a lower bound on the ring radius. It is easy to see that for either
positive or negative cosmological constant this radius is always
outside the black hole horizon at $\rho_{bh}$. Below this $R_l$ the
attraction of the black hole is too strong for the ring to resist its
pull. 

\medskip

We believe it is possible to extend this analysis to the case where
the central black hole is rotating, but this lies outside the scope of
this paper.

\section{No supersymmetric black rings in Anti-deSitter}
\label{sec:nosusy}

Supersymmetric black rings exist in five-dimensional ungauged
supergravities, so it is natural to look for their counterparts in
gauged supergravities. Imagine that such solutions exist with horizon
topology $S^1\times S^2$, and that the radius of the $S^2$ can be made
arbitrarily smaller than the $S^1$-radius and the cosmological length,
\ie that thin small supersymmetric black rings exist. As a limit we should
recover a straight black string solution of the ungauged theory.
Moreover, if the circular ring solution is supersymmetric, the
limiting straight string should be supersymmetric as well. In general the
straight solution will have no fewer (super)symmetries than the circular
one, since some of the symmetries of the straight string geometry may
be broken when bent into a circle\footnote{In fact, if we curve a
straight supersymmetric string, in general it will not remain
supersymmetric.}. 

Thus AdS supersymmetric black rings, if they exist
and admit a thin small string limit, should be amenable to our approximate
methods.
Let us consider, for simplicity, minimal five-dimensional gauged supergravity--- the
case with arbitrary vector multiplets presents no qualitative
differences. Our arguments above imply that a thin supersymmetric black
ring, when expanded in powers of $r_0/R$, $r_0/L$, should  be described to zeroth
order by the supersymmetric black string of ungauged supergravity
\cite{Bena:2004wv}
\beqa\label{susystring}
ds^2&=&-f^{2}(dt+A_z dz)^2+f^{-1}(dr^2+r^2d\Omega_3^2)\,,\\
f^{-1}&=&1+\frac{Q}{r}+\frac{q^2}{4r^2}\,,\qquad
A_z=-\left(\frac{3q}{2r}+\frac{3qQ}{4r^2}+\frac{q^3}{8r^3}\right)\,\nonumber
\eeqa
(here $r_0=q/2$), which indeed arises as a limit of supersymmetric black
rings in the ungauged theory \cite{Elvang:2004rt}.
To obtain a circular ring by curving this string in AdS,
we proceed as in
section~\ref{sec:thinrings} and identify the distributional source that
reproduces the linearized field \eqref{susystring} at
distances $r\gg r_0$. This is easy to obtain, since terms $\propto q^2,
qQ$ in the metric are non-linear and can be dropped. We find
\beqa
T_{tt}= \frac{3Q}{16\pi G}\;\delta^{(3)}(r) \,,\qquad
T_{tz}= \frac{3q}{32\pi G}\;\delta^{(3)}(r)\,,\qquad
T_{zz}= 0\,.
\eeqa
Crucially, observe that the tension vanishes. We believe this to be a
generic consequence (`cancellation of forces') of supersymmetry.

The next step is solving the linearized field equations when this source is
placed along a circle of radius $R$ in AdS. But we have seen in
eq.~\eqref{equil} that a consistent coupling of this {\em circular} source to
gravity in AdS requires non-vanishing pressure, which the supersymmetric
string cannot provide.\footnote{Note that the stress-energy tensor of the
electromagnetic field will only enter at quadratic order in $q$ and $Q$.
Also, besides conservation of energy-momentum one must impose
conservation of charge, but this is trivially satisfied for a uniform
circular profile.} Thus there cannot be supersymmetric black rings in
AdS that admit a limit to \eqref{susystring}. 

More precisely, if we insist on bending the string \eqref{susystring}
into a circle, the tension required in eq.~\eqref{equil} will appear in
the form of conical singularities on the plane of the ring. These
conical singularities arose in the analysis of
\cite{Kunduri:2006uh,Kunduri:2007qy} as
the obstruction preventing the existence of near-horizon geometries of
supersymmetric AdS black rings. Our argument above confirms these
results and gives a simple physical origin to the conical
singularities.

\section{Final remarks}
\label{sec:conclusions}

Our analysis of black rings in the presence of a cosmological constant
demonstrates the power of the approach initiated in
\cite{Emparan:2007wm} for the investigation of black holes with
non-spherical topologies in higher-dimensional theories. A simple,
straightforward technique provides non-trivial information in cases
where the construction of exact solutions has been unsuccessful. Of all
the solutions we have constructed approximately, the one that would seem
more likely to admit a closed exact analytic solution is the static
black ring in deSitter in five dimensions, generalizing the fairly
simple metric for a vacuum static black ring \cite{Emparan:2001wk}. The
problem in this case seems to lie more in the form of the metric far
from the horizon than close to it.

Although we have not done it here, one could very easily include
additional spins for the black ring, corresponding to rotation of the
$S^{d-3}$. This is simply done by considering a rotating boosted black
string, and bending it to form a ring. The equilibrium condition
\eqref{equil} is not affected by this rotation in transverse directions, so
it is straightforward to compute all physical magnitudes. 

We believe that the same approach should be useful to study black
rings in many other backgrounds. For instance, it may be interesting to
apply it to black rings in G\"odel spacetimes and study how the ring
behaves as its radius becomes larger than the scale at which closed
timelike curves appear. 

Some of our conclusions may also be applicable to other extended objects
(p-branes). Our argument for the absence of supersymmetric black rings
in AdS is based on the simple observation that the AdS potential
requires pressure along the p-brane in order to achieve mechanical
balance when the p-brane is curved into some contractible cycle. But
supersymmetry is presumably incompatible with this pressure, and so,
according to this argument, in AdS we should not expect supersymmetric
branes whose worldvolumes have contractible cycles --- that is, unless
some other force is present to achieve the balance without breaking all
supersymmetries.

We have discussed our black rings as living in AdS space, but in the
context of string theory it is more natural to consider them as
solutions in a larger space, say, AdS$_5\times S^5$, with the radius of
$S^5$ being given by $L$. In this case, our solutions describe black
rings which are completely smeared over the $S^5$, \ie their horizons
are $S^1\times S^2\times S^5$. Our approximations require that the $S^2$
be much smaller than the $S^5$, in which case a Gregory-Laflamme
instability towards localization in $S^5$ is expected. This is in
addition to the GL-instability along the $S^1$. If the final fate of
both instabilities is the fragmentation of the black ring, then we would
end up with ten-dimensional black holes, with $S^8$ horizons, flying
apart (and possibly merging eventually into a single black hole with
$S^8$ horizon). For black rings beyond our approximations, depending on
the possible regimes of existence (discussed in
section~\ref{sec:mergers}), they may localize in $S^5$ without
fragmenting along $S^1$, or remain extended along $S^5$ while
fragmenting along $S^1$, giving $S^3\times S^5$ horizons. It is also
possible that large stable black rings exist, in particular if they also
possess charges or dipoles.

An interesting approach to the physics of new black holes in AdS is
based on their dual fluid-dynamical description. We have already
discussed how the analysis in
\cite{Bhattacharyya:2007vs,Bhattacharyya:2008ji} has not found any
evidence for large AdS black rings nor for any phases of large pinched
black holes (which are in any case unlikely, since pinched black holes
should branch off from small AdS black holes with $r_+<L$). On the other
hand, plasma rings and plasma balls, including pinched plasma balls,
have turned up in the study of hydrodynamic solutions in a theory with a
confining vacuum \cite{Lahiri:2007ae,Bhardwaj:2008if}. The gravitational
duals of such solutions, which are not any of the black holes and black
rings of this paper, are in many respects more similar to asymptotically
flat black holes. Perhaps an extension of the hydrodynamic methods will
allow to describe some of the solutions we have constructed, and, maybe,
clarify some of the puzzles that we have left unsolved.

\section*{Acknowledgments}

We thank Veronika Hubeny and Tom\'as Ort{\'\i}n for discussions. This
work was supported in part by DURSI 2005 SGR 00082, MEC FPA
2004-04582-C02 and FPA-2007-66665-C02, and the European Community FP6
program MRTN-CT-2004-005104. MJR was supported in part by an FI
scholarship from Generalitat de Catalunya.

\section*{Appendices}

\begin{appendix}

{\bf NOTE:} in appendices \ref{app:adapted} and \ref{app:bendstring} we denote
the total number of spacetime dimensions as
\beq
d=n+4\nonumber
\eeq
in order to facilitate comparison with the results of \cite{Emparan:2007wm}.

\section{Adapted coordinates and near-zone `potential'}
\label{app:adapted}

\subsection{Anti-deSitter}

Close to the circle at $\rho=R$, $\Theta=0$ in \eqref{globalads} we
can choose a set of coordinates $(t,r,\theta,\Omega_i,z)$ in AdS which
are adapted to ring-like objects, in the sense that the
radial coordinate $r$ measures transverse distance away from the circle
at $r=0$ and surfaces of constant $r$ have ring-like topology $S^1\times
S^{n+1}$. The geometry close to the ring can then be studied by expanding the metric
in powers of $r/R$.

To obtain these coordinates, first observe that the constant $\tau$
hypersurfaces of
(\ref{globalads}) are
($d-1$)-dimensional hyperbolic spaces, as is obvious introducing the new
radial coordinate $\hat \rho$,
\eq
\rho=\frac{\hat \rho}{1-\hat \rho^2/(4L^2)}\,,
\eeq
which brings AdS$_d$ into homogeneous (spatially conformally flat) coordinates
\eq
ds^2=-\lp\frac{1+\hat \rho^2/(4L^2)}{1-\hat \rho^2/(4L^2)}\rp^2 d\tau^2
+\frac{1}{\lp 1-\hat \rho^2/(4L^2)\rp^2}\lp d\hat \rho^2+
\hat \rho^2d\Theta^2+\hat \rho^2\sin^2\Theta\,d\Omega_{n}^2
+\hat\rho^2\cos^2\Theta d\psi^2\rp\,.
\eeq
Note that $\hat \rho$ ranges from zero at the origin to $\hat \rho=2L$ at
spatial infinity of AdS. The ring will be located at $\Theta=0$, $\hat \rho
=\hat R$,
such that
\beq
R=\frac{\hat R}{1-\hat R^2/(4L^2)}\,.
\eeq
Define coordinates $(\hat r,\theta)$ as
\beq
\hat r \sin\theta =\hat \rho \sin\Theta\,,\qquad \hat r \cos\theta =\hat \rho
\cos\Theta -\hat R\,.
\label{hatrtheta}\eeq
The ring source lies at $\hat r=0$, and spatial surfaces at
$\hat r=\rm{const.}$
have topology $S^1\times S^{n+1}$. We may now expand the metric near $\hat
r=0$ in powers of $\hat r/\hat R$, but before doing that it is
convenient to make a few additional changes. Notice first that to
zero-th order in $\hat r/\hat R$ we have
\beq
ds^2=-\lp 1+\sr^2 \rp d\tau^2
+\frac{R^2}{\hat R^2}
\left[ d\hat r^2+\hat r^2\lp d\theta^2+\sin^2\theta\,d\Omega^2_{n}\rp\right]
+R^2 d\psi^2
+\mc{O}(\hat r/R)
\eeq
Thus, orthonormal coordinates $(t, z)$ in which $t$ is proper time
and $z$ proper length along
the string worldheet directions, can be defined as in
\eqref{orthocoord}. We also rescale the radial coordinate to
measure proper radius $r$,
\beq
r= \frac{R}{\hat R}\,\hat r\,.
\eeq
In terms of these, the metric including the first corrections in
$ r/R$, is
\beqa
ds^2&=&- \left(1+\frac{\sr^2}{\sqrt{1+\sr^2}}
\frac{2r\cos\theta}{R}+\mc{O}(r^2/R^2)\right)dt^2\nonumber\\
&&+ \left(1+\left(\sqrt{1+\sr^2}-1\right)\frac{2r\cos\theta}{R}
+\mc{O}(r^2/R^2)\right)
\left(dr^2+r^2 d\theta^2+r^2\sin^2\theta\,d\Omega^2_{n}\right)
\nonumber\\
&&+\left(1+\sqrt{1+\sr^2}\;\frac{2r\cos\theta}{R}
+\mc{O}(r^2/R^2)\right)d z^2
\,.
\eeqa

This metric is of the generic form
\beqa\label{backpot}
ds^2&=&-\left(1+C_t
\frac{2r\cos\theta}{R}+\mc{O}(r^2/R^2)\right)dt^2\nonumber\\
&&+ \left(1+C_r\frac{2r\cos\theta}{R}
+\mc{O}(r^2/R^2)\right)
\left(dr^2+r^2 d\theta^2+r^2\sin^2\theta\,d\Omega^2_{n}\right)
\nonumber\\
&&+\left(1+C_z\frac{2r\cos\theta}{R}
+\mc{O}(r^2/R^2)\right)d z^2
\,.
\label{genadapted}\eeqa
with constant $C_{t,r,z}$, which describes a generic class of
backgrounds in which the straight line at $r=0$ is deformed into a large
circle, whose extrinsic curvature radius in the plane $\theta=0$
at constant $t$ is $R/C_z$.
The deformations in the metric only involve a dipole perturbation ($\propto
\cos\theta$) of the $S^{n+1}$.
A gauge transformation of the form
\beq
r\to r +a\frac{r^2\cos\theta}{R}+\mc{O}(R^{-2})\,,\qquad
\theta\to \theta+a\frac{r\sin\theta}{R}+\mc{O}(R^{-2})\,,
\eeq
does not modify $g_{tt}$ nor $g_{tz}$ nor does it introduce crossed
terms $g_{r\theta}$, but it shifts
\beq
C_r \to C_r +2 a\,.
\eeq
We can use this gauge freedom to obtain a more convenient radial
coordinate. If we demand that $\Box r^{-n}=0$, to order
$\mc{O}(R^{-2})$, so $r$ labels scalar equipotential surfaces, then we
must adjust $a$ to have
\beq\label{harmgauge}
C_r=-\frac{C_t+C_z}{n}\,.
\eeq
We make this choice to finally obtain
\beqa\label{orho2}
ds^2&=&- \left(1+\frac{\sr^2}{\sqrt{1+\sr^2}}
\frac{2r}{R}\cos\theta\right)d t^2\nonumber\\
&&+ \left(1-
\frac{1+2\sr^2}{n\sqrt{1+\sr^2}}\,
\frac{2r}{R}\cos\theta
\right)
\left(dr^2+r^2 d\theta^2+r^2\sin^2\theta\,d\Omega^2_{n}\right)
\nonumber\\
&&+\left(1+\sqrt{1+\sr^2}\;\frac{2r}{R}\cos\theta
\right)d z^2+\mc{O}(r^2/R^2)
\,.
\eeqa
If we keep only terms of order $\mc{O}(r^0/R^0)$ we obviously recover
flat space where $r=0$ is a straight line along $z$. The corrections
included in \eqref{orho2} capture the bending of this line into an arc
of large radius $R$. But the Riemann tensor of this metric vanishes up
to terms $\mc{O}(r^2/R^2)$, so to this order the metric \eqref{orho2}
(and \eqref{backpot}) is in fact still {\it flat} space. Even if it was
obtained starting from AdS space, the cosmological constant
$\Lambda\propto -L^{-2}$ does not contribute terms of lower order than
$(r/L)^2=\sr^2 (r/R)^2$, and so it can only
appear in the metric through the combination $\sr^2=R^2/L^2$.

The terms of order $r/R$ in eq.~\eqref{orho2} can be regarded as a
`gravitational potential': a black string placed in this field will
be bent into an arc of large radius $R$. If the thickness of the black
string is $r_0$, the approximation requires that the correction terms in
\eqref{orho2} are small for $r\sim r_0$, \ie
\beq\label{smallro}
r_0\ll \frac{R}{\sqrt{1+\sr^2}}\,.
\eeq	
Up to factors of order one, this implies the validity range
\eqref{smallroRL}.

\subsection{General static spherically symmetric backgrounds}

Here we show that the near-zone static metric \eqref{backpot} includes
all static spherically symmetric
spacetimes. 
These always admit a spatially
conformally flat static metric, 
\eq
ds^2=-f(\hat\rho) d\tau^2
+g(\hat\rho)\lp d\hat \rho^2+
\hat \rho^2d\Theta^2+\hat \rho^2\sin^2\Theta\,d\Omega_{n}^2
+\hat\rho^2\cos^2\Theta\,d\psi^2\rp\,.
\label{genmetric}
\eeq
To obtain the geometry of the region near the ring located at
$\Theta=0$, $\hat\rho=\hat R$, we perform the coordinate transformation
(\ref{hatrtheta}), and define the coordinates $t$ and $z$ that measure
the proper time and proper length along the string world-sheet
directions, and the proper radius $r$ by
\eq
t=\sqrt{f(\hat R)}\,\tau,\qquad
z=\hat R\sqrt{g(\hat R)}\,\psi,\qquad
r=\sqrt{g(\hat R)}\,\hat r.
\eeq
This casts the metric into adapted coordinates for the circular string,
which takes, up to order $r/R$, the form (\ref{genadapted}) with
coefficients
\eq
C_t=\left.\frac{(\ln f)'}{2\sqrt{g}}\right|_{\hat\rho=\hat R} R\,,\qquad
C_r=\left.\frac{(\ln g)'}{2\sqrt{g}}\right|_{\hat\rho=\hat R} R\,,\qquad
C_z=\left.\frac{2+\hat R(\ln g)'}{2\hat R\sqrt{g}}\right|_{\hat\rho=\hat R}
R\,.
\eeq
Here, $R$ is a constant with the dimensions of length that sets the
scale of the perturbations. We can conveniently take it as 
the proper circumference radius of the circle
\beq
R=\frac{\Delta z}{2\pi}=\hat
R\sqrt{g(\hat R)}\,.
\eeq

The particular class of metrics \eqref{globalads} is related to the
generic metric (\ref{genmetric}) through a
redefinition of the radial coordinate $\rho=\rho(\hat\rho)$ such that 
\eq
\frac{d\rho}{d\hat\rho}=\frac\rho{\hat\rho}\sqrt{V(\rho)}\,.
\eeq
Using this relation, and $R=\rho(\hat R)$, it is easy to show that the coefficients $C_{t,r,z}$
simplify to
\eq\label{CV}
C_t=\frac{R V'(R)}{2\sqrt{V(R)}}\,,\qquad
C_r=\sqrt{V(R)}-1\,,\qquad
C_z=\sqrt{V(R)}\,,
\eeq
with $z=R\psi$ and the rescaling encoding the redshift of the black string given
by
$t=\sqrt{V(R)}\,\tau$.
The constant $C_r$ can be
gauge-transformed to the value \eqref{harmgauge} in the manner described
above. 

\section{Bending the black string}
\label{app:bendstring}

In this appendix we extend the analysis of \cite{Emparan:2007wm} to
describe the perturbations of a boosted black string induced by placing
it in a `background potential' of the generic form \eqref{backpot} (with
gauge choice \eqref{harmgauge}). It seems likely that any static
background that possesses a hypersurface-orthogonal spacelike Killing
vector should reduce, near an orbit of the isometry, to this
form\footnote{For stationary backgrounds, one can choose corotating
coordinates that eliminate the term $g_{tz}$ to zero-th order in $r/R$,
but in general not to higher orders. Thus one must supplement
\eqref{backpot} with a term $g_{tz}\propto r\cos\theta/R$.}.

Our main aim is to show that the horizon of the black string remains
regular after we bend it to a circular shape. The analysis is simply an
extension of that of \cite{Emparan:2007wm}, so our description will be
very succinct.

Typically, in a matched asymptotic expansion with two separate scales
$r_0\ll R$ we would begin by solving first the equations in the far-zone
$r\gg r_0$ and then use this solution in the intermediate-zone $r_0\ll
r\ll R$ in order to provide boundary conditions for the solution in the
near-zone $r\ll R$.

In the case of AdS we have found it quite difficult to solve the
far-zone equations. However, we have managed to fully determine the
solution in the intermediate-zone, up to gauge transformations. This
solution, which we obtain in the next subsection, is enough to
provide the boundary conditions for the perturbations of the black
string. Analyzing the problem in this manner has the advantage that we
can deal with the larger class of backgrounds \eqref{backpot}. Recall
that
the particular case of AdS
corresponds to
\beq\label{ctcz}
C_t=\frac{\sr^2}{\sqrt{1+\sr^2}}\,,\qquad C_z=\sqrt{1+\sr^2}\,,
\eeq
while the results of \cite{Emparan:2007wm} are recovered with $C_t=0$, $C_z=1$.

\subsection{Intermediate-zone analysis}

The intermediate zone is defined by
\beq
r_0\ll r\ll \min{(R,L)}\,,
\eeq
or, in more generality for a background of the form \eqref{backpot},
\beq
r_0\ll r\ll R\min{(C_t^{-1},C_z^{-1})}\,.
\eeq

In this case the ring will be a small perturbation around
\eqref{backpot} and we can solve the equations in the linearized
approximation. Like in \cite{Emparan:2007wm}, we model the ring by a generic
string-like distributional stress-energy tensor of the form
\beqa
T_{tt}&=&\frac{n(n+2)}{n+1}\;\mu\;\frac{r_0^{n}}{16\pi G}
\;\delta^{(n+2)}(r) \nonumber\,,\\
T_{tz}&=&n\,p\,\frac{r_0^{n}}{16\pi G}\; \delta^{(n+2)}(r)\,,\label{Ttz}\\
T_{zz}&=&\frac{n(n+2)}{n+1}\;\tau\;\frac{r_0^{n}}{16\pi G}\; \nonumber
\delta^{(n+2)}(r)\,,
\eeqa
where $\mu$, $p$, $\tau$, are dimensionless quantities characterizing
the mass and momentum densities and the pressure along the string. For
the particular case of a boosted black string (cf.\
eq.~\eqref{tensorcomponents}),
\beqa\label{mutaup}
\frac{n(n+2)}{n+1}\mu&=&n\,\cosh^2\alpha +1\,,\label{mubeta}\nonumber\\
\frac{n(n+2)}{n+1}\tau&=&n\,\sinh^2\alpha -1\,,\label{taubeta}\label{pbeta}\\
p&=&\cosh\alpha\sinh\alpha\,. \nonumber
\eeqa

Conservation of this stress-energy tensor, $\nabla_\mu T^{\mu\nu}=0$, in
the background
\eqref{backpot} implies that
\beq\label{taumu}
\tau =\frac{C_t}{C_z}\mu\,,
\eeq
which is actually equivalent to \eqref{equil}. We shall impose this condition
henceforth --- if at this stage we did not, it would be later enforced
on us as a regularity condition on the perturbed geometry like in
\cite{Emparan:2007wm}.
When applied to \eqref{mutaup}, eq.~\eqref{taumu} fixes the equilibrium
value of the boost to
\beq\label{genequil}
\sinh^2\alpha=\frac{C_z+(n+1)C_t}{n(C_z-C_t)}\,.
\eeq
This reproduces \eqref{boost} for the Anti-deSitter background
\eqref{ctcz}, and \eqref{boostV} for the background \eqref{globalads}
with generic function $V(\rho)$.

Observe that, according to the observation made after eq.~\eqref{orho2}, the
equations that we must solve in this region are the vacuum Einstein
equations $R_{\mu\nu}=0$.
Following the same steps as in \cite{Emparan:2007wm}, we find that the regular
solution to the linearized Einstein's equations is
\beqa
g_{tt}&=&-\left(1+C_t\frac{2r\cos\theta}{R}\right)
   \left(1- \left(\mu+\frac{\tau}{n+1}\right)\frac{r_0^n}{r^n}\right)\,,\nonumber\\
g_{tz}&=&-p\frac{r_0^{n}}{r^{n}}
    \left(1+\frac{C_t+C_z}{2}\frac{2r\cos\theta}{R}\right)\,,\nonumber\\
g_{zz}&=&\left(1+C_z\frac{2r\cos\theta}{R}\right)
   \left(1-\left(\tau+\frac{\mu}{n+1}\right)\frac{r_0^n}{r^n}\right)\,,\\
g_{rr}&=&\left(1-\frac{C_t+C_z}{n}\frac{2r\cos\theta}{R}\right)
   \left(1+\frac{\mu-\tau}{n+1}\frac{r_0^n}{r^n}
   -k(n-1)\frac{r_0^n}{r^n}\frac{2r\cos\theta}{R}\right)\,,\nonumber\\
g_{ij}&=&\hat g_{ij}
   \left(1-\frac{C_t+C_z}{n}\frac{2r\cos\theta}{R}\right)
   \left(1+\frac{\mu-\tau}{n+1}\frac{r_0^n}{r^n}
   +k\frac{r_0^n}{r^n}\frac{2r\cos\theta}{R}\right)\,,\nonumber
\eeqa
where in the angular part we have the factor
\beq\label{hatg}
\hat g_{ij}dx^i dx^j= r^2(d\theta^2 +\sin^2\theta d\Omega^2_{n})\,.
\eeq
There is an undetermined constant $k$ from the solution to the
homogeneous differential equations. This was set to zero in \cite{Emparan:2007wm} using
an argument that does not apply here, but in any case it is just a gauge
perturbation: to the required perturbation order, we can set it to zero
by a coordinate transformation of the form
\beq
r\to r-k\frac{n-1}{n-2}\frac{r_0^{n}}{r^{n-2}}\frac{\cos\theta}{R}\,,\qquad
\theta\to \theta +\frac{k}{n-2}\frac{r_0^{n}}{r^{n-1}}\frac{\sin\theta}{R}\,.
\eeq

In the following, for simplicity we set $k=0$\footnote{We would actually
need the far-zone solution to check that this choice is consistent with
the asymptotic boundary
conditions, which may fix partially this gauge freedom.
We could in any case leave $k$ free and our
conclusions would not be modified.} and assume that the string is a
boosted
black string
with energy-momentum parameters \eqref{pbeta}. For the purpose of
analyzing the
near-horizon perturbations it is convenient to
pass to yet another gauge, $r\to r-\frac{r_0^n}{2n r^{n-1}}$, in
which the solution takes the form
\beqa\label{asympform}
g_{tt}&=&-1+c_\alpha^2\frac{r_0^n}{r^n}
-C_t\frac{2r\cos\theta}{R}\left[1-\frac{r_0^n}{r^n}c_\alpha^2
\left(1+\frac{1}{2n c_\alpha^2}\right)\right]
\,,\nonumber\\
g_{tz}&=&-c_\alpha s_\alpha \frac{r_0^{n}}{r^{n}}
    \left(1+\frac{C_t+C_z}{2}\frac{2r\cos\theta}{R}\right)\,,\nonumber\\
g_{zz}&=&1+s_\alpha^2\frac{r_0^n}{r^n}
+C_z\frac{2r\cos\theta}{R}\left[1+\frac{r_0^n}{r^n}s_\alpha^2
\left(1-\frac{1}{2n s_\alpha^2}\right)\right]
\,,\\
g_{rr}&=&1+\frac{r_0^n}{r^n}-\frac{C_t+C_z}{n}\frac{2r\cos\theta}{R}
\left[1+\frac{r_0^n}{r^n}\left(1-\frac{1}{2n}\right)\right]\,,\nonumber\\
g_{ij}&=&\hat g_{ij}\left[
   1-\frac{C_t+C_z}{n}\frac{2r\cos\theta}{R}
\left(1-\frac{1}{2n}\frac{r_0^n}{r^n}\right)
\right]\,,\nonumber
\eeqa
where we abbreviate
\beq
c_\alpha=\cosh\alpha\,,\qquad s_\alpha=\sinh\alpha\,.
\eeq

\subsection{Perturbations near the horizon}

We now study the perturbations of the boosted black string, considering
that $r\ll R$ but without
assuming $r\gg r_0$. This generalizes the
analysis of \cite{Emparan:2007wm} to allow for a general value of the
boost $\alpha$. The effect of the background potential \eqref{backpot}
will only enter when we impose boundary conditions. Only then will we
need to fix the boost to the value \eqref{genequil}.

The same arguments as in \cite{Emparan:2007wm} allow us to
reduce the perturbations to the form
\begin{equation}
\label{gtta}
g_{tt} = - 1 +c_\alpha^2 \frac{r_0^{n}}{r^{n}} +
\frac{\cos \theta}{R} a(r)\,,
\end{equation}
\begin{equation}
g_{tz} = -c_\alpha s_\alpha \left[ \frac{r_0^{n}}{r^{n}} +
\frac{\cos \theta}{R} b(r) \right]\,,
\end{equation}
\begin{equation}
g_{zz} = 1 +s_\alpha^2\frac{r_0^{n}}{r^{n}} + \frac{\cos \theta}{R}
c(r)\,,
\end{equation}
\begin{equation}\label{grr}
g_{rr} = \left( 1- \frac{r_0^{n}}{r^{n}} \right)^{-1} \left[ 1
+ \frac{\cos \theta}{R} f(r) \right]\,,
\end{equation}
\begin{equation}
\label{gijg}
g_{ij} =  \hat g_{ij}\left[ 1
+ \frac{\cos \theta}{R} g(r) \right]\,,
\end{equation}
where $\hat g_{ij}$ is the metric \eqref{hatg}.
This form is convenient since it
fixes the location of the horizon at $r=r_0$. There remains a gauge
freedom
\beqa
r\to r+ \gamma(r)\,\frac{r_0}{R}\,\cos\theta\,,\qquad
\theta\to\theta + \frac{r_0}{R}\,\sin\theta\int^r
dr'\frac{\gamma(r')}{{r'}^2\left(1-
\frac{r_0^{n}}{{r'}^{n}}\right)}\,,
\eeqa
that we deal with by working with invariant variables
\beqa\label{Ainvt}
{\sf A}(r)&=&a(r)-\frac{c_\alpha^2}{s_\alpha^2}\,c(r)\,,\\
{\sf B}(r)&=&b(r)-\frac{1}{s_\alpha^2} \,c(r)\,,\label{Binvt}\\
{\sf F}(r)&=&f(r)+\frac{2}{n\,s_\alpha^2} r_0\left(\frac{r^{n+1}}{r_0^{n+1}}\;
c(r)\right)'
-\frac{1 }{\left(1-
\frac{r_0^{n}}{r^{n}}\right)\,s_\alpha^2}\;c(r)\,,\label{Finvt}\\
{\sf G}'(r)&=&g'(r)+\frac{2}{n\,s_\alpha^2}
 \frac{r_0}{r}\left(\frac{r^{n+1}}{r_0^{n+1}}\;c(r)\right)'
+\frac{2}{n s_\alpha^2\,r\left(1- \frac{r_0^{n}}{r^{n}}\right)}
\;c(r)\label{Gpinvt}\,.
\eeqa

The Einstein equations can now be shown to imply that, for any boost
$\alpha$, the functions $\sf A$ and $\sf
B$ satisfy the same fourth-order differential equation as derived in
\cite{Emparan:2007wm}. As a consequence we can immediately write their solution as
\beqa\label{AandB}
{\sf A}(r)&=& A_1\; u_1(r)  + A_2\; u_2(r)\,,\nonumber\\
{\sf B}(r)&=& B_1\; u_1(r)  + B_2\; u_2(r)\,,
\eeqa
with
\beqa
u_1(r)&=&\ _2F_1\left(-\frac{1}{n},-\frac{n+1}{n};1;
    1-\frac{r_0^n}{r^n}\right)r\,,\nonumber\\
    \
u_2(r)&=&\ _2F_1\left(-\frac{1}{n},\frac{n-1}{n};1;
    1-\frac{r_0^n}{r^n}\right)\frac{r_0^n}{r^{n-1}}\,.
\eeqa
We have discarded solutions that blow up at $r=r_0$.

Note that ${\sf A}(r)$ and ${\sf B}(r)$ are not independent but one can
be obtained from the other through an equation that in general differs
slightly from the one in \cite{Emparan:2007wm}. However, we shall not need this
equation since the boundary behavior we obtained in the previous
subsection is enough to fix all the integration constants $A_i$ and
$B_i$. The remaining
functions ${\sf F}(r)$
and ${\sf G}(r)$ are obtained from ${\sf A}(r)$ and ${\sf B}(r)$
as
\beqa
{\sf F}(r)&=&-\frac{ 2 (1+c_{\alpha}^2 n) 
r^{2 n}-(4+c_{\alpha}^2 (3n-2)) r^n r_0^n-2 s_{\alpha}^2 r_0^{2 n}}
{n \left(r^n-r_0^n\right)r_0^{n}} \,
{\sf A}-\frac{2 r \left((1+c_{\alpha}^2 n) 
r^n+s_{\alpha}^2 r_0^n\right)}{n (n+1)r_0^{n}}\,{\sf A'}\nn\\
&&+\frac{2 s_{\alpha}^2 c_{\alpha}^2  
\left(n r^n \left(2 r^n-3 r_0^n\right)+2 r_0^n \left(r^n-r_0^n\right)\right)}
{n \left(r^n-r_0^n\right)r_0^{n}} \,{\sf B}+\frac{4 s_{\alpha}^2 
c_{\alpha}^2 r  \left(n r^n+r_0^n\right)}{n (n+1)r_0^{n}} \,{\sf B'}\,,
\eeqa
\beqa
{\sf G}'(r)&=&-\frac{2((1+n c_{\alpha}^2 )r^n-(1+(n-1)c_{\alpha}^2 )r_0^n)r^{n-1} }
{n(r^n-r_0^n)r_0^{n}} \,{\sf A}
-\frac{2(1+n c_{\alpha}^2 )r^n+(n+2)s_{\alpha}^2 r_0^n}{n(n+1)r_0^n} \,{\sf A'}\nonumber\\
&&+\frac{4 s_{\alpha}^2 c_{\alpha}^2 (nr^n -(n-1)r_0^n)r^{n-1}}{n(r^n-r_0^n)r_0^n} \,
{\sf B}+\frac{2 s_{\alpha}^2 c_{\alpha}^2 (2nr^n+(n+2)r_0^n)}{n(n+1)r_0^n} \,{\sf B'}\,.
\eeqa
\paragraph{Asymptotic boundary conditions.}

We fix the integration constants $A_i$, $B_i$ by demanding that the
solution asymptotes to \eqref{asympform} as $r\to\infty$, \ie that
\beq\label{asympAB}
{\sf A}(r)=2r\left({\sf a}_0+{\sf a}_1\left(\frac{r_0}{r}\right)^n+O(r^{-n-
2})\right)\,,\quad
{\sf B}(r)=2r\left({\sf b}_0+{\sf b}_1\left(\frac{r_0}{r}\right)^n+O(r^{-n-
2})\right)\,
\eeq
with
\beqa
{\sf a}_0&=&-\left(C_t+\frac{c_\alpha^2}{s_\alpha^2}C_z\right)\,,\qquad
{\sf a}_1=c_\alpha^2(C_t-C_z)+\frac{1}{2n}
\left(C_t+\frac{c_\alpha^2}{s_\alpha^2}C_z\right)\,,\nonumber\\
{\sf b}_0&=&-\frac{C_z}{s_\alpha^2}\,,\qquad
{\sf b}_1=\frac{C_t-C_z}{2}+\frac{C_z}{2n s_\alpha^2}\,.
\eeqa
We assume that the boost takes on the equilibrium value \eqref{genequil}.
Expanding \eqref{AandB} at large $r$ and comparing to \eqref{asympAB} determines
\beq\label{A1A2}
A_1=\frac{2}{\pi}\frac{n+1}{n^3}\,\Gamma\left(\frac{1}{n}\right)^2
\Gamma\left(-\frac{n+2}{n}\right)\sin\left(\frac{2\pi}{n}\right){\sf
a}_0\,,\qquad A_2=2A_1\left(\frac{1}{n}+\frac{n+2}{n+1}\frac{{\sf a}_1}{{\sf
a}_0}\right)\,
\eeq
and
\beqa\label{B1B2}
B_1=A_1\frac{{\sf b}_0}{{\sf a}_0}\,,\qquad
B_2=2A_1\left(\frac{1}{n}\frac{{\sf b}_0}{{\sf a}_0}
+\frac{n+2}{n+1}\frac{{\sf b}_1}{{\sf
a}_0}\right)\,.
\eeqa

\paragraph{Regularity of the horizon.}

The solution is now fully determined, up to gauge transformations, and
it only remains to check that the horizon stays regular. This requires that
the angular velocity and surface gravity are constant quantities on the
horizon and do not depend on the polar angle $\theta$.

Constancy of the horizon angular velocity is easily seen to be equivalent to
\beq
{\sf A}(r_0)=(c_\alpha^2+s_\alpha^2){\sf B}(r_0)
\eeq
which translates into $A_1+A_2=(c_\alpha^2+s_\alpha^2)(B_1+B_2)$. This
{\em is} satisfied by \eqref{A1A2}, \eqref{B1B2}. Additionally, one must
partially fix the gauge freedom by requiring that the gauge-variant
function $c(r)$ at the horizon be $c(r_0)=-c_\alpha^2
s_\alpha^2(B_1+B_2)r_0$.

Uniformity of the surface gravity requires in turn
\beq
a'(r_0)-2 s_\alpha^2\, b'(r_0)
+\frac{s_\alpha^2}{c_\alpha^2}\,c'(r_0)+
\frac{n}{c_\alpha^2 r_0}
f(r_0)=0\,.
\eeq
Despite appearances, this condition is 
gauge-invariant for regular gauge transformations. Using the explicit
form of the solution derived above, this equation is seen to be
identically satisfied.

\medskip

Thus we conclude that we can construct black rings that are regular on
and outside the horizon by placing a boosted black string in any
spacetime that, locally near the ring, takes the form \eqref{backpot}.

\section{Phase space of rotating AdS black holes}
\label{app:MPAdS}

Here we discuss the phase space of the rotating AdS
black hole solution of \cite{Gibbons:2004uw} with more than one angular
momentum (the case of a single angular momentum is discussed in the main
text). Their mass, angular momenta, area and surface gravity, as computed in
\cite{Gibbons:2004ai}, are
\begin{equation}
M=\frac{m\,{\Omega}_{d-2}}{4\pi\prod_j\Xi_j}
\left(\sum_{i=1}^N\frac1{\Xi_i}-\frac{1-\epsilon}{2}\right),\qquad
J_i=\frac{m\,{\Omega}_{d-2}a_i}{4\pi\Xi_i\prod_j\Xi_j}\,,
\end{equation}
\begin{equation}
{\mathcal A}=\frac{\Omega_{d-2}}{r_+^{1-\epsilon}}
\prod_{i}\frac{r_+^2+a_i^2}{\Xi_i}\,,\qquad
\kappa=r_+\lp1+\frac{r_+^2}{L^2}\rp\lp\sum_i\frac1{r_+^2+a_i^2}+\frac\epsilon{2r_+^2}\rp-\frac1{r_+}\,,
\end{equation}
where $N=\left[\frac{d-1}2\right]$ is the maximal number of independent
angular momenta, $a_i$ are the $N$ angular velocity parameters, $m$ is
the mass parameter, $\epsilon=(d-1)\mathrm{mod}\,2$, $\Xi_i=1-{a_i^2}/{L^2}$ 
and $r_+$ is the horizon position, given by the largest root of the equation
\eq
r^{\epsilon-2}(1+r^2/L^2)\prod_{i}(r^2+a_i^2)-2\,m=0.
\label{horizon}\eeq
Note that this equation is the same as the equation for the horizon of
the asymptotically flat MP black hole in $d+2$ dimensions,
where the additional rotation is $a_{N+1}=L$ and mass parameter is
$\mu=2L^2m$ \cite{Emparan:2008eg}. Hence, the root structure and the
horizons of the rotating AdS$_d$ black hole can be inferred from the
MP$_{d+2}$ solution; in particular, for odd $d$, a horizon always
exists provided that any two of the spin parameters vanish, while for
even $d$, its existence is guaranteed if any one of the spins vanishes.
Therefore, under this assumption, an ultraspinning limit can be achieved
for all but two (one) of the $a_i\rightarrow L$ in odd (even)
dimensions.

\begin{figure}[t!]
\begin{center}
\includegraphics[width=7cm]{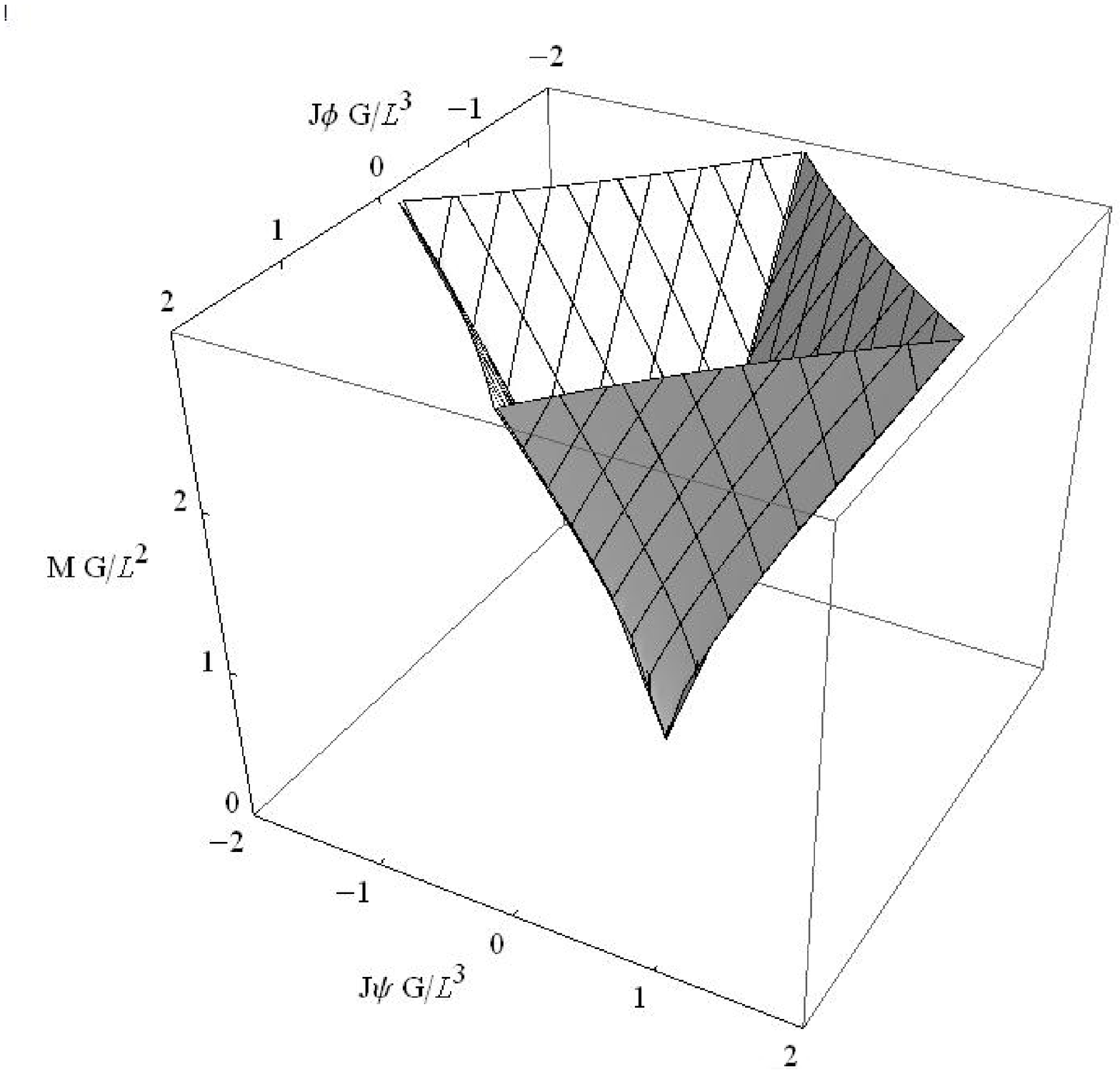}
\hspace{1cm}
\includegraphics[width=7cm]{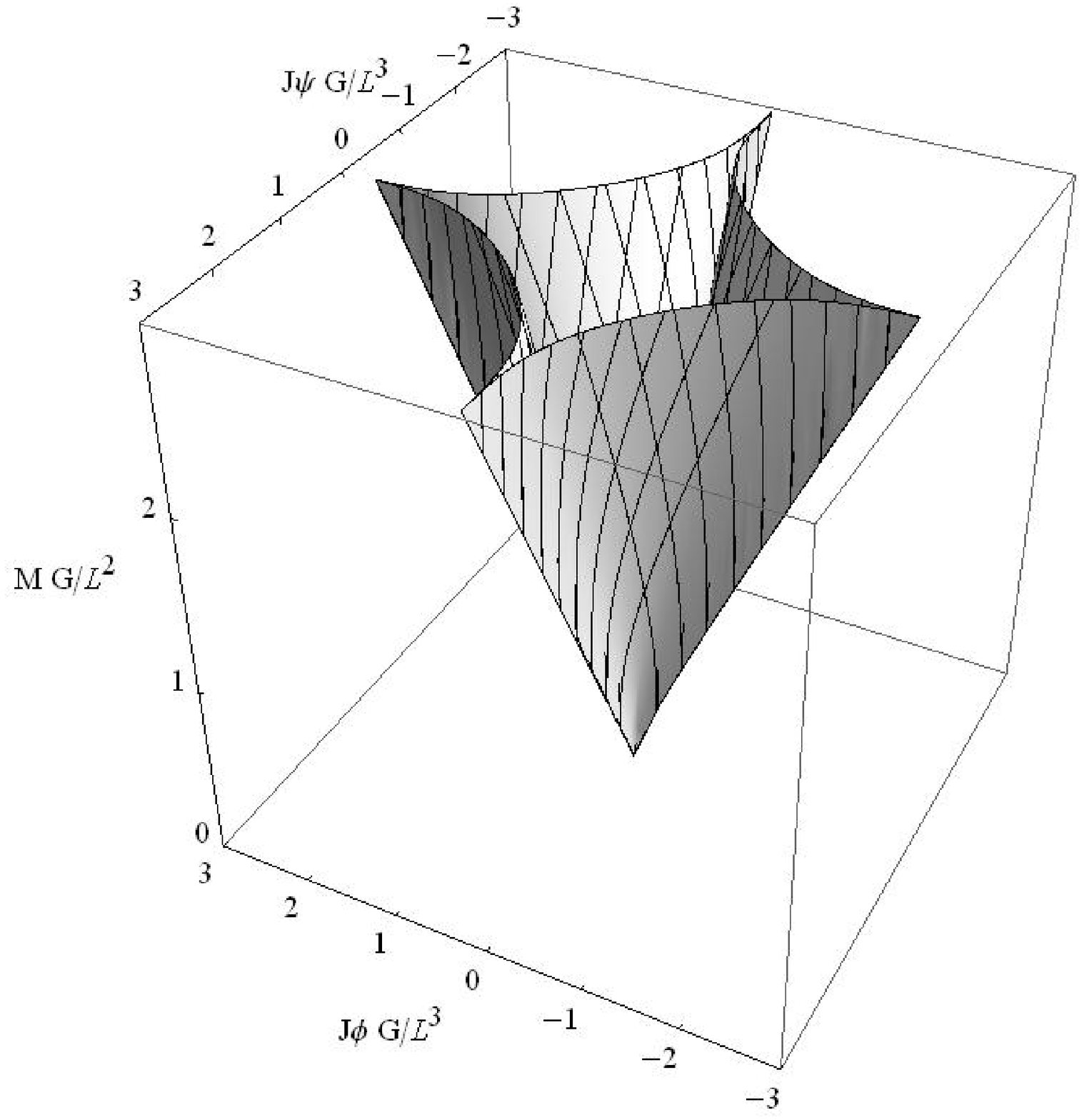}
\end{center}
\begin{picture}(0,0)(0,0)
\put(40,2){\footnotesize (a)}
\put(118,2){\footnotesize (b)}
\end{picture}
\caption{
\small Boundaries of the phase space $(\sj_{\phi},\sj_{\psi},\sm)$
covered by doubly spinning
rotating AdS black holes in five (a) and six (b) dimensions. The mass
increases along the vertical axis. The surfaces correspond to extremal,
zero-temperature black holes, except at the edges where they become
naked singularities. The interior of the pyramids is filled by
non-extremal black holes.
}\label{fig:AdS5/AdS6}
\end{figure}

\begin{figure}
\begin{center}
{\includegraphics[height=4.5cm]{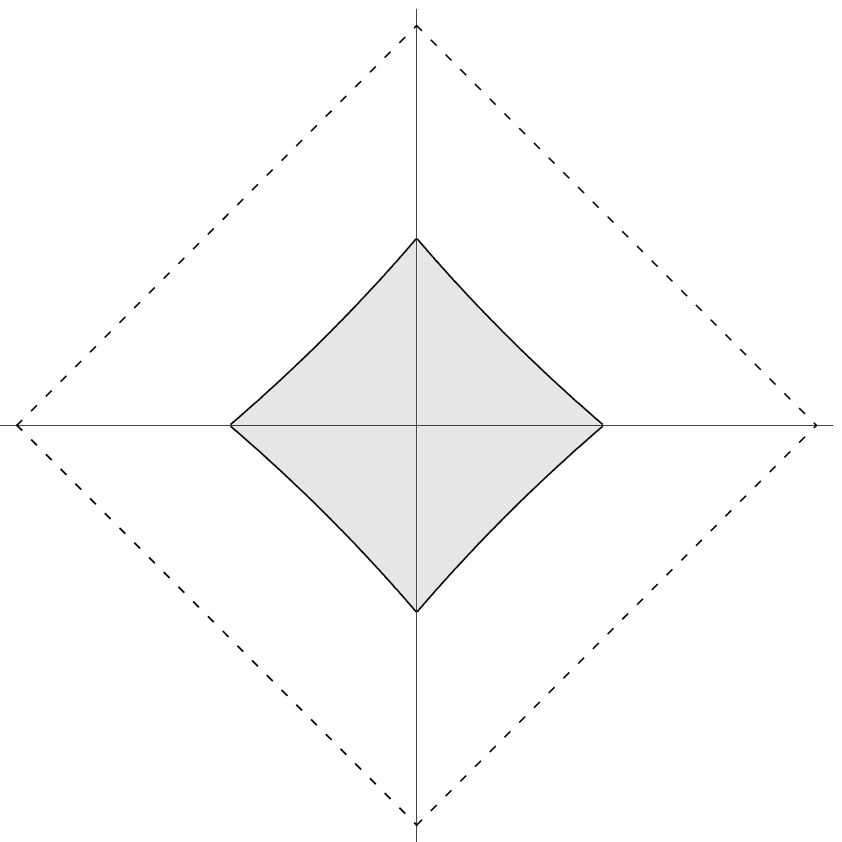}}
\hspace{1.4cm}
{\includegraphics[height=4.5cm]{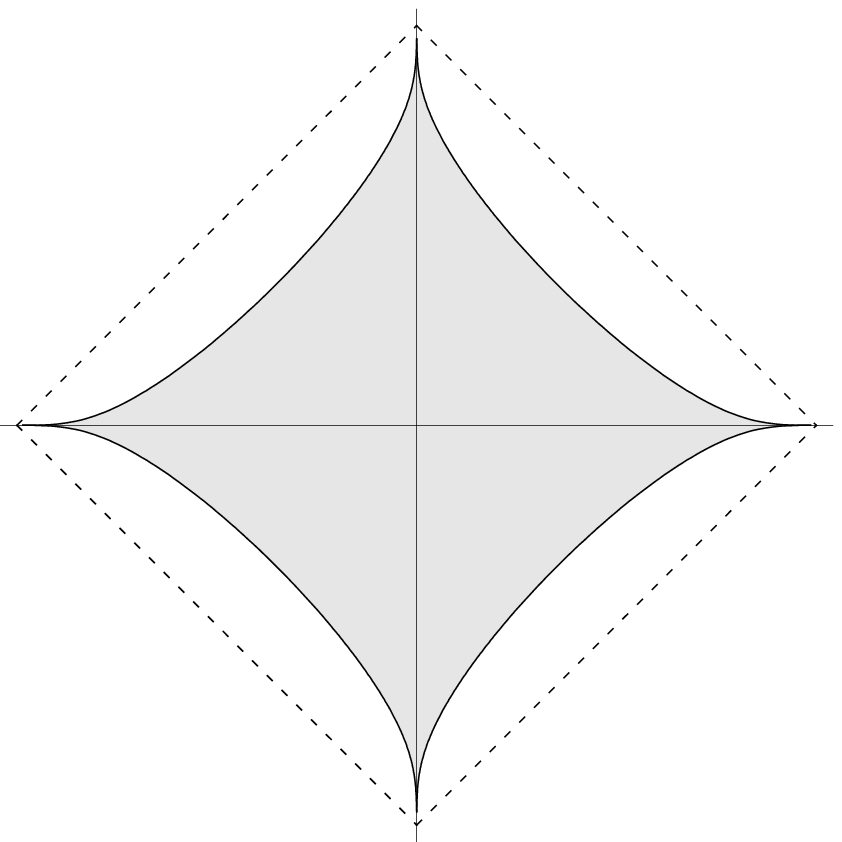}}
\end{center}
\begin{picture}(0,0)(0,0)%
\put(52,57){\small $\sj_{\phi}$}
\put(78,32){\small $\sj_{\psi}$}
\put(113,57){\small $\sj_{\phi}$}
\put(140,32){\small $\sj_{\psi}$}
\put(50,3){\footnotesize (a)}
\put(112,3){\footnotesize (b)}
\end{picture}
\caption{\small Cuts at constant $\sm$ of the surfaces of
fig.~\ref{fig:AdS5/AdS6}. Black hole solutions exist for angular momenta
in the shaded regions. The
dashed lines denote the BPS limit $\sum_{i=1}^N\left|\sj_i\right|=\sm $.
In six dimensions this is saturated only at the
corners. These diagrams can be compared to fig.~2 of \cite{Emparan:2008eg}.
In more than six dimensions, we expect the diagrams to resemble the ones
of the asymptotically flat case, see \cite{Emparan:2008eg}, only
`compressed' to lie within the `BPS diamond'.
}
\label{fig:AdS5/AdS6BPS}
\end{figure}

The phase space $(\sm,\sj_{\phi},\sj_{\psi})$ covered by the rotating AdS
black holes is shown in figure \ref{fig:AdS5/AdS6} for the five and six
dimensional cases, filling the interior of the pyramids (the
five-dimensional diagram was already presented in \cite{Emparan:2008eg}). The
extremal solutions lie at the faces of the pyramid.
Fig.~\ref{fig:AdS5/AdS6BPS} presents cuts of these surfaces at a
constant value of the mass.

We can show that these black holes comply with the ``BPS bound''
\beq
ML\geq \sum_{i=1}^N \left|J_i\right|
\label{posenergy}
\eeq
proven in \cite{Chrusciel:2006zs}. To this effect, write
\begin{equation}
\frac{1}{ML}\sum_{i=1}^{N}\left|J_i\right|=
\frac{\sum_i\frac{|a_i|}{L\Xi_i}}{\sum_j\frac{1}{\Xi_j}-\frac{1-\epsilon}2}\,.
\label{JE}
\end{equation}
If $d$ is even, $\epsilon=1$, and since $|a_i|/L\leq1$ we have
\begin{equation}
\sum_i\frac{|a_i|}{L\Xi_i}\leq\sum_j\frac{1}{\Xi_j}\,,
\end{equation}
proving equation (\ref{posenergy}).
If $d$ is odd, $\epsilon=0$ and we can rewrite equation~(\ref{JE}) as
\begin{equation}
\frac{1}{ML}\sum_{i=1}^{N}\left|J_i\right|=
\frac{\sum_i\frac{|a_i|/L}{\Xi_i}}{\sum_j\frac{1-\Xi_j/2N}{\Xi_j}}\,.
\end{equation}
The polynomial
\begin{equation}
P_i(a_i)=1-\frac{\Xi_i}{2N}-\frac1L |a_i|=
\frac{a_i^2}{2NL^2}-\frac{|a_i|}{L}+1-\frac1{2N}
\end{equation}
admits always, as smaller positive root, $a_i=L$, and is therefore
positive for $|a_i|<L$. Hence,
\begin{equation}
\frac{|a_i|/L}{\Xi_i}\leq\frac{1-\Xi_j/2N}{\Xi_j}\,,
\end{equation}
and (\ref{posenergy}) follows.

It is easy to show that this bound can only be saturated in the
ultra-spinning regime, in which one or more spin parameters tend to $L$.
However, it can never be saturated when all the angular momenta are non-zero. Indeed, suppose $n$
spin parameters approach the ultraspinning limit. To keep the mass
finite, we need to scale the parameters as
\eq
\Xi_{\al=1\ldots n}=\xi_\al\nu\,,\qquad m=\mu\nu^{n+1},
\eeq
where $\nu\rightarrow0$ in the ultraspinning limit, while keeping
$\xi_1,\ldots,\xi_n$ and $\mu$ constant. As we observed previously, this
limit is allowed provided any one (two) of the $a_i$ vanish in even
(odd) dimensions. Then the root $r_+$ of equation (\ref{horizon}) tends
to zero, while the mass and angular momenta reach the values
\eq
M=\frac{\mu\Omega_{d-2}}{4\pi\Pi_\al\xi_\al\Pi_I\Xi_I}\sum_\al\frac1{\xi_\al}\,,\qquad
J_\al=\frac{\mu\Omega_{d-2}}{4\pi\xi_\al\Pi_\beta\xi_\beta\Pi_I\Xi_I}\,,\qquad
J_I=0\,,
\eeq
(with $\al,\beta=1\ldots n$ running on the spin parameters that tend to
$L$ and $I=n+1,\ldots,N$ denoting the others) and saturate the BPS bound
(\ref{posenergy}). However, these black holes
are not extremal, since the surface gravity diverges like
$\kappa\rightarrow(2k+\epsilon-2)/2r_+$, where $k$ is the number of
vanishing spin parameters. In this limit the area of the horizon
decreases to zero like
\beq
\sa \propto \sm^{\frac{2k+\epsilon-1}{2k+\epsilon-2}}
\left(1-\frac{\sj}{\sm}\right)^{\frac{2k+n+\epsilon-1}
{2k+\epsilon-2}}\left(1+\mc{O}(\sm-\sj)\right)\,,
\eeq
and the limiting geometry describes a black membrane with horizon
topology $\mathbb{R}^{2n}\times S^{d-2(n+1)}$. Since in this limit the
black holes are pancaked out along the planes of rotation, it is
reasonable to presume that they will develop a Gregory-Laflamme type of
instability.

\end{appendix}

\end{document}